\font\llbf=cmbx10 scaled\magstep2
\font\lbf=cmbx10 scaled\magstep1
\def\sq{\hfill \vbox {\hrule width3.2mm \hbox {\vrule height3mm \hskip 2.9mm
                      \vrule height3mm} \hrule width3.2mm }}
\def\ni{\noindent}

\def\edth{\hskip 3pt {}^{\prime } \kern -6pt \partial }
\def\thorn{ { \rceil \kern -6pt \supset }}

\def\ua{\underline a \,}
\def\ub{\underline b \,}

\def\uA{\underline A \,}

\def\bM{\bf M\,}
\def\bN{\bf N\,}
\def\bK{\bf K\,}
\def\bL{\bf L\,}
\def\bR{\bf R\,}
\def\bS{\bf S\,}

\def\bi{\bf i\,}
\def\bj{\bf j\,}
\def\bk{\bf k\,}

\baselineskip 14pt plus 2pt

%%%%%%%%%%%%%%%%%%%%%%%%%%%%%%%%%%%%%%%%%%%%%%%%%%%%%%%%%%%%%%%%%%%%%%
%%%%%%%%%%%%%%%%%%%%%%%%%%%%%%%%%%%%%%%%%%%%%%%%%%%%%%%%%%%%%%%%%%%%%%

\ni
{\llbf Quasi-local holography and quasi-local mass of classical fields 
in Minkowski spacetime}

\bigskip
\ni
{\bf L\'aszl\'o B. Szabados}

\ni
Research Institute for Particle and Nuclear Physics

\ni
H-1525 Budapest 114, P.O.Box 49, Hungary

\ni
E-mail: lbszab@rmki.kfki.hu

\bigskip
\bigskip
\ni
Dedicated to the memory of Zolt\'an Perj\'es (1943-2004), the 
originator of the relativity research in Hungary.

\bigskip
\ni
The 2-surface characterization of special classical radiative Higgs-, 
Yang--Mills and linear zero-rest-mass fields with any spin is 
investigated. We determine all the zero quasi-local mass Higgs- and 
Yang--Mills field configurations with compact semisimple gauge groups, 
and show that they are plane waves (provided the Higgs field is 
massless and linear) and appropriate generalizations of plane waves 
(`Yang--Mills {\it pp}-waves'), respectively. A tensor field 
(generalizing the energy-momentum tensor for the Maxwell field and 
of the Bel--Robinson tensor for the linearized gravitational field) 
is found by means of which the {\it pp}-wave nature of the solutions 
of the linear zero-rest-mass field equations with any spin can be 
characterized equivalently. It is shown that these radiative 
Yang--Mills and linear zero-rest-mass fields, given on a finite 
globally hyperbolic domain $D$, are determined completely by certain 
unconstrained data set on a {\it closed spacelike 2-surface}, the `edge 
of $D$'. These pure radiative solutions are shown to determine a 
dense subset in the set of solutions of various (Yang--Mills and 
linear zero-rest-mass) field equations. Thus for these fields 
some `classical quasi-local holography' holds. 

\bigskip
\bigskip

\ni
{\lbf 1 Introduction}
\medskip
\ni
By combining the basic principles of quantum theory and general 
relativity 't Hooft [1] argued that the number of the physical 
degrees of freedom of a localized system surrounded by a closed 
2-surface ${\cal S}$ should be bounded from above by $C\,{\rm Area}
({\cal S})/4L_P^2$, where $C$ is constant of order one and $L^2_P:=
\hbar G$, the Planck area. This bound appears to be quite robust, 
and a similar bound can be derived for the number $N$ of quantum 
particles in a box bounded by a spherical ${\cal S}$ of radius $R$ 
by requiring that the sum of the zero-point energy of the quantum 
particles be less than or equal to the Schwarzschild mass 
${1\over2G}R$ corresponding to $R$. Indeed, by the localization 
of the particles in the box the zero-point energy of one particle 
is $\hbar\omega=C\,{\pi\hbar\over R}$ for some constant $C$ of order 
one, and hence $2\pi^2C\,N\leq{\rm Area}({\cal S})/4L_P^2$. (This 
inequality can also be interpreted as an upper bound for the 
number $N$ of the possible quantum states of a single localized 
particle with energy not greater than the Schwarzschild mass.) 
Moreover, combining Bekenstein's entropy-to-energy bound [2] for 
weakly gravitating systems, $S/E\leq k\sqrt{\pi{\rm Area}({\cal S})}/
\hbar$, and the requirement that $E$ should not be greater than the 
corresponding Schwarzschild mass, Bousso [3] got $S/k\leq{\rm Area}
({\cal S})/4L_P^2$. (Here $k$ is Boltzmann's constant.) By the 
statistical interpretation of the entropy this is essentially the 
previous inequality. Therefore, the physical degrees of freedom in 
a domain of the 3-space grow with the {\it area} of the boundary of 
the domain, rather than with its volume. This observation lead 't 
Hooft to the so-called holographic hypothesis [1]: The physical 
state of a system bounded by ${\cal S}$ must be possible to be 
characterized by degrees of freedom on ${\cal S}$ with information 
density not greater than one bit per Planck area. 

Susskind [4] implemented this principle mathematically by assuming 
the existence of a map from the 3-space to the 2-surface using light 
rays arriving orthogonally at ${\cal S}$, the so-called `screen'. 
However, the space in general relativity is a spacelike hypersurface 
$\Sigma$, and if the screen is its boundary, ${\cal S}:=\partial\Sigma$, 
then $\Sigma$ may be mapped by null geodesics to the {\it world tube} 
of ${\cal S}$, which is a timelike submanifold, but certainly not to 
the screen itself. Nevertheless, in a spacetime admitting a preferred 
timelike vector field (e.g. a hypersurface orthogonal timelike Killing 
field) the null geodesics can be projected to $\Sigma$, and these 
could be used to define the map $\Sigma\rightarrow{\cal S}$. Corley and 
Jacobson refined this idea of Susskind, and they clarified the global 
properties of such maps [5]. The recent formulations of the 
holographic principle (and of the covariant entropy bounds) are also 
based on this view [3,6]. 

Here we suggest another mathematical formulation of the holographic 
principle. The overall picture is based on that we preferred in 
looking at the various quasi-local energy-momentum expressions [7]: 
${\cal S}$ is the common boundary of all the spacelike hypersurfaces 
that are Cauchy surfaces for the domain of dependence $D(\Sigma)$ of 
$\Sigma$. Thus, ${\cal S}$ can be considered as being associated with 
$D(\Sigma)$ rather than only to an individual Cauchy surface: ${\cal 
S}$ is the `edge' of $D(\Sigma)$. Then, according to the holographic 
principle, the state of the matter+gravity system on the {\it four 
dimensional} $D(\Sigma)$ should be able to be characterized by 
appropriate data on the {\it two dimensional} screen ${\cal S}$. This 
formulation of the holographic hypothesis is, in some sense, 
causally complement to the previous ones [3-6] based on the use of 
null geodesics. Indeed, by $I^\pm[{\cal S}]\cap D(\Sigma)=\emptyset$ 
and $\partial I^\pm[{\cal S}]\cap{\rm int}\, D(\Sigma)=\emptyset$ 
the screen ${\cal S}$ is causally disjoint from the interior of 
$D(\Sigma)$. This would be a field theoretical formulation of the 
principle, using certain `elementary solutions' of the field 
equations on $D(\Sigma)$ to `scan' the space $\Sigma$, in contrast to 
the previous, basically (general relativistic) geometric optical 
approach. 

The motivation of this picture came from the results of the 
investigations of the Dougan--Mason quasi-local energy-momentum and 
mass [8]: The energy-momentum, associated with a spacelike 2-surface 
${\cal S}$ with spherical topology, is vanishing iff $D(\Sigma)$ is 
flat; and the energy-momentum is null (i.e. the mass is zero) iff 
$D(\Sigma)$ has a {\it pp}-wave geometry and the matter is pure 
radiation [9,10]. Furthermore, it has been demonstrated that in 
the latter case the geometry (and also the matter fields in the 
form of linear massless scalar or Maxwell fields) on $D(\Sigma)$ 
are completely determined by the spinor geometry of (and, in the 
presence of matter fields, by an additional complex function on) 
${\cal S}$ [11]. Thus the state of the special radiative gravitational 
field inside ${\cal S}$, specified by the vanishing of a basic 
quasi-local observable, the mass, can be fully characterized by 
an appropriate data set on ${\cal S}$. 

The aim of the present paper is to investigate the validity of the 
present formulation of the holographic principle for certain classical 
matter fields in Minkowski spacetime. In particular, we discuss the 
connection between the following three concepts: 

\item{1.} The quasi-local mass of the matter fields, associated with 
  a closed spacelike 2-surface ${\cal S}$ in Minkowski spacetime, 
\item{2.} The pure radiative solutions of specific matter field 
  equations (Higgs, Yang--Mills, linear zero-rest-mass fields with 
  any spin) on the globally hyperbolic domain $D(\Sigma)$ for which 
  the common boundary of the Cauchy surfaces $\Sigma$ is just ${\cal 
   S}$, 
\item{3.} The complete characterization of these radiative solutions 
  in terms of data specified on the 2-surface ${\cal S}$ rather than 
  the familiar Cauchy data on $\Sigma$. 

\noindent
Since the 2-surface ${\cal S}$ will be assumed to be a topological 
2-sphere (i.e. the `screen' is finite, two dimensional, and it is 
at finite distance from the physical system), the holography that 
we study here will be called `classical quasi-local holography'. 
We show that the vanishing of the quasi-local mass is equivalent 
to the pure radiative nature of the matter fields, these special 
radiative solutions can be characterized by data sets on the screen, 
and that they determine a dense subset in the space of regular 
matter fields. Thus, ironically enough, while the holographic 
hypothesis came from the combination of the basic principles of the 
gravitational and quantum physics, some form of the holographic 
hypothesis could be shown to work at the classical level in the 
absence of gravitation. Though the results depend on the flatness 
of spacetime, many of them can be generalized for curved spacetimes 
e.g. by considering the Dougan--Mason quasi-local mass of the 
matter+gravity system in Section 3. (Indeed, the Dougan--Mason 
energy-momentum is based on the 2-surface integral of the 
Nester-Witten 2-form, the 3-surface integral of whose exterior 
derivative could be written as the sum of the energy-momentum for 
the matter fields [via Einstein's equations] and of the so-called 
Sparling 3-form. The latter can be interpreted as the energy-momentum 
of the gravitational `field' itself.) On the other hand, it is not 
quite clear how all the results can be generalized. Nevertheless, 
we hope that the present investigations could be instructive for an 
analysis in more general spacetimes. 

In the second section we recall some facts about the geometry of 
closed spacelike 2-surfaces, compact spacelike hypersurfaces with 
boundary, and the `finite' globally hyperbolic domains. The role 
of this section is to fix the notations and present the geometric 
background to make the present paper essentially self-contained. 
Next, in Section 3, we briefly recall the notion of quasi-local 
mass for general matter fields and study its properties. First we 
show that the quasi-local mass is zero precisely when the matter 
fields on $D(\Sigma)$ are of pure radiative in the sense that 
$T_{ab}L^b=0$ for some {\it constant} null vector field $L^a$ on 
$D(\Sigma)$, provided $T_{ab}$ satisfies the dominant energy 
condition. Then we consider special matter fields. According to 
the standard model the basic building blocks of Nature are the 
fermions and the gauge fields, represented by genuine spinor and 
(non-Abelian) Yang--Mills fields, respectively, and by an additional 
scalar field, called the Higgs field. Thus we determine 
all the solutions of the Higgs and Yang--Mills field equations 
(with any compact semisimple gauge groups) for which the 
quasi-local mass is zero. These are all plane waves (provided the 
Higgs field is massless and linear, i.e. non-self-interacting) and 
appropriate generalizations of plane waves (`Yang--Mills {\it 
pp}-waves'), respectively. 

Since the spinor fields in general do not satisfy the dominant 
energy condition, the general results above cannot be applied to 
characterize the spinor {\it pp}-waves by the vanishing of the 
quasi-local mass. However, a totally symmetric, trace-free tensor 
field $T_{a_1...a_{2s}}$ is found, which is a natural generalization 
of the energy-momentum tensor of the Maxwell field and of the 
Bel--Robinson tensor for the linearized gravitational field, by 
means of which the {\it pp}-wave character of the solutions of the 
linear zero-rest-mass field equations with spin $s$ on $D(\Sigma)$ 
can be characterized equivalently. All these solutions are 
represented by a constant spinor field $O^A$ and a complex function 
$\phi(u,\zeta)$ on $D(\Sigma)$, where $O^A$ is the spinor constituent 
of the null wave vector $L^a=O^A\bar O^{A'}$ (and hence we call it 
the `wave spinor') and $\phi(u,\zeta)$ is $C^2$ in the coordinate $u$, 
labeling the wave-fronts, and complex analytic in $\zeta$, the 
anti-holomorphic coordinate in the intersection of the wave-fronts 
and $\Sigma$. These results are presented in Section 4. 

In Section 5 we determine those data sets on ${\cal S}$ by means of 
which the {\it pp}-waves on $D(\Sigma)$ can be fully characterized: 
Provided a convexity condition for the 2-surface ${\cal S}$ is 
satisfied, it is shown that a pair $(O^A,\phi)$ on $D(\Sigma)$ is 
completely determined by a constant spinor field $O^A$ on the 
2-surface ${\cal S}$ and two real valued functions $(f(u,w),g(u))$ of 
two and one variables, respectively, where $w$ is a coordinate on 
the intersection of ${\cal S}$ and the wave-fronts. The data set 
$(O^A,f(u,w),g(u))$ on ${\cal S}$ is unconstrained, and the field 
$\phi$ at any point $p\in D(\Sigma)$ is expressed as an appropriate 
contour integral on ${\cal S}$. Finally, we show that the {\it 
pp}-wave solutions of the Weyl equation with a given, fixed 
wave spinor $O^A$ form so large a space $\Phi(O^A)$ that any spinor 
field (with arbitrary spin $s$) on $\Sigma$ can be uniformly 
approximated with arbitrary accuracy by the elements of the tensor 
product $\Phi(O^A)\otimes\bar\Phi(O^A)$, where $\bar\Phi(O^A)$ is the 
complex conjugate of $\Phi(O^A)$. Therefore, this tensor product is 
a dense subspace in the space of all continuous spinor fields. The 
consequences of this result are also discussed. The Stone--Weierstrass 
theorem that we use in the proof is stated in the Appendix. 

Throughout this paper the abstract index formalism of [12] will be 
used, and only the underlined and boldface indices take numerical 
values. The signature of the spacetime metric is $-2$, and we use 
units in which the speed of light is $c=1$. 
\bigskip
\bigskip

\ni
{\lbf 2 Finite Cauchy developments with constant null vector field}
\medskip
\ni
Although in the present paper primarily we are interested in the 
quasi-local characterization of fields in Minkowski spacetime, in 
the present section we do not restrict our attention to flat 
spacetimes. It is enough to be only a {\it pp}-wave spacetime, and, 
for later use, in the subsequent discussion we assume only this. 

\medskip
\ni
{$\bullet$ The geometry of closed spacelike 2-surfaces}

\ni
Let ${\cal S}$ be a closed, orientable spacelike 2-surface in $M$. 
Let $t^a$ and $v^a$ be a future pointing timelike and an outward 
directed spacelike unit normal of ${\cal S}$, respectively, such 
that $t^av_a=0$. Then $\Pi^a_b:=\delta^a_b-t^at_b+v^av_b$ is the 
orthogonal projection to ${\cal S}$ and $q_{ab}:=\Pi^c_a\Pi^d_bg_{cd}$ 
is the induced metric on ${\cal S}$. Then the corresponding induced 
area 2-form is $\varepsilon_{cd}:=t^av^b\varepsilon_{abcd}$. The 
extrinsic curvatures corresponding to the normals $t^a$ and $v^a$ 
are defined, respectively, by $\tau_{ab}:=\Pi^c_a\Pi^d_b\nabla_ct_d$ 
and $\nu_{ab}:=\Pi^c_a\Pi^d_b\nabla_cv_d$. The expansion tensor of the 
standard outgoing and ingoing null normals $l^a:=t^a+v^a$ and $n^a:=
{1\over2}(t^a-v^a)$, respectively, are $\theta_{ab}=\tau_{ab}+\nu
_{ab}$ and $\theta'_{ab}={1\over2}(\tau_{ab}-\nu_{ab})$. Then the 
2-surface ${\cal S}$ will be called convex [13,11] if ${\cal S}$ is 
homeomorphic to $S^2$ and the outgoing null normals are expanding 
($q^{ab}\theta_{ab}>0$), the ingoing null normals are contracting 
($q^{ab}\theta'_{ab}<0$) and, in addition, $2\det\Vert\theta^a{}_b\Vert=
(\theta_{ab}\theta_{cd}-\theta_{ac}\theta_{bd})q^{ab}q^{cd}>0$ and $2\det
\Vert\theta'^a{}_b\Vert=(\theta'_{ab}\theta'_{cd}-\theta'_{ac}\theta'_{bd})
q^{ab}q^{cd}>0$ also hold. Note that because of the scaling freedom 
$l^a\mapsto\alpha l^a$, $n^a\mapsto\alpha^{-1}n^a$ the normals $l^a$ and 
$n^a$ (or $t^a$ and $v^a$) are not uniquely determined by ${\cal S}$, 
and the expansion tensors $\theta_{ab}$ and $\theta'_{ab}$ depend on 
the choice for the normals $l^a$ and $n^a$, the convexity of ${\cal 
S}$ is well defined. 

If $L^a$ is any nowhere vanishing null vector field on ${\cal S}$, 
then $z_a:=\Pi^b_aL_b$ is vanishing at a point of ${\cal S}$ 
precisely when $L^a$ is orthogonal to ${\cal S}$. If $L_a$ is 
constant on ${\cal S}$ in the sense that $\Pi^c_a\nabla_cL_b=0$, 
then by the definitions we have $\delta_az_b=\nu_{ab}v^cL_c-\tau_{ab}
t^cL_c$, where $\delta_a$ is the intrinsic Levi-Civita derivative 
operator on $({\cal S},q_{ab})$. This implies that $\delta_{[a}z
_{b]}=0$, i.e. $z_a$ is a closed 1-form on ${\cal S}$. If ${\cal S}$ 
is homeomorphic to $S^2$, then by $H^1(S^2)=0$ it follows the 
existence of a smooth function $u:{\cal S}\rightarrow{\bf R}$ such 
that $z_a=\delta_au$. Clearly, $s^a:=\varepsilon^{ab}z_b$ is a vector 
field tangent to the $u={\rm const}$ level sets of ${\cal S}$. 

\medskip
\ni
{$\bullet$ The geometry of compact spacelike hypersurfaces with 
boundary} 

\ni
Let $\Sigma$ be a smooth, compact orientable spacelike hypersurface 
with smooth 2-boundary ${\cal S}$. Let $t^a$ be the future pointing 
unit timelike normal of $\Sigma$, and $P^a_b:=\delta^a_b-t^at_b$, the 
orthogonal projection to $\Sigma$. Then the induced metric is $h_{ab}
:=P^c_aP^d_bg_{cd}$ and the extrinsic curvature of $\Sigma$ in $M$ 
is defined by $\chi_{ab}:=P^c_aP^d_b\nabla_ct_d$ (and hence on its 
boundary $\tau_{ab}=\Pi^c_a\Pi^d_b\chi_{cd}$ holds). If $L^a$ is any 
nowhere zero null vector field, then since $\Sigma$ is spacelike, 
$Z^a:=P^a_bL^b$ is nowhere vanishing on $\Sigma$. Hence it can be 
decomposed as $L^a=Z^a+\Vert Z\Vert t^a$, where $\Vert Z\Vert^2:=-h
_{ab}Z^aZ^b$ is the (everywhere positive) norm of $Z^a$. If $L^a$ is 
null and constant on $\Sigma$ in the sense that $P^b_a\nabla_bL^c=0$, 
then $D_aZ_b=-\Vert Z\Vert\chi_{ab}$, where $D_a$ is the intrinsic 
Levi-Civita derivative operator on $\Sigma$. This implies that 
$D_{[a}Z_{b]}=0$, and hence $Z_a=D_au$ for some (in general only {\it 
locally} defined) function $u$ on $\Sigma$. If, in addition, $L^a$ 
is constant in the spacetime in the sense that $\nabla_bL^a=0$, then 
$L_a=\nabla_au$ for some real valued function $u$. We assume that the 
domain of dependence $D(\Sigma)$ is contained in the domain of this 
function $u$, and hence $u$ on $\Sigma$ is globally defined. Since 
$Z_a$ is nowhere vanishing, $u$ defines a foliation of $\Sigma$ by 
its level surfaces $S_u:=\{p\in\Sigma\vert u(p)=u\}$. These are 
smooth spacelike 2-surfaces. The induced Riemannian metric determines 
a unique complex structure on the surfaces $S_u$; i.e. they are 
Riemann surfaces. 

Let $\{E^a_1,E^a_2,E^a_3\}$ be an orthonormal frame field on $\Sigma$ 
such that $E^a_3:={1\over\Vert Z\Vert}Z^a$, i.e. $L^a=\Vert Z\Vert(t^a
+E^a_3)$. Then $E^a_3$ is globally well defined on $\Sigma$, and the 
(locally defined) complex null vectors $\sqrt2M^a:=E^a_1-{\rm i}E^a_2$ 
and $\bar M^a$ are (1,0) and (0,1) type vectors in the complex 
structure of the Riemann surfaces $S_u$, respectively. The vectors 
$L^a$, $M^a$ and $\bar M^a$ can be completed to be a Newman--Penrose 
complex null tetrad on $\Sigma$ by introducing the future pointing 
null vector field $N^a$ such that $N^a$ be orthogonal to the 
2-surfaces $S_u$ and normalized according to $L^aN_a=1$. Explicitly, 
it is given by $2\Vert Z\Vert N^a=t^a-E^a_3$. 
The lapse function of the foliation $\{S_u\}$ of $\Sigma$ is just 
$\Vert Z\Vert^{-1}$. Thus the `acceleration' of $\{S_u\}$ in $\Sigma$ 
is $E^a_3D_aE_{3b}=(\delta^a_b+\Vert Z\Vert^{-2}Z^aZ_b)\Vert Z\Vert D_a
({1\over\Vert Z\Vert})$, while its extrinsic curvature in $\Sigma$ is 
${}^uk_{ab}:=(\delta^e_a+E^e_3E_{3a})(\delta^f_b+E^f_3E_{3b})D_e(-E
_{3f})=(\delta^e_a+E^e_3E_{3a})(\delta^f_b+E^f_3E_{3b})\chi_{ef}$. (It 
is $-E^a_3$ along that $u$ is {\it increasing}, thus it seems natural 
to choose $-E^a_3$ to be the normal of $S_u$ in $\Sigma$.) Then by the 
Gauss equation the integrability condition of $L^aL_a=0$, $\nabla_a
L_b=0$ implies that the scalar curvature ${}^uR$ of the surfaces $S_u$ 
is zero; i.e. $S_u$ are locally flat Riemann geometries. 

Without further restrictions the topology of the Riemann surfaces 
may be very complicated. If, however, ${\cal S}$ is convex in the 
sense above, then by Proposition 4.6 of [11] 
\smallskip
\item{i.}  $\Sigma$ is homeomorphic to the closed three-ball $B^3$; 
\item{ii.} each Riemann surface $S_u$ is homeomorphic to ${\bf R}^2$; 
          and 
\item{iii.} each $S_u$ intersects the 2-boundary ${\cal S}$ in a 
          single smooth closed curve $\gamma_u:={\cal S}\cap S_u$, 
          $u$ has a single minimum $p_-$ and a maximum point $p_+$ 
          on ${\cal S}$, and hence $\{z^a,s^a\}$ on ${\cal S}-\{p_+,
            p_-\}$ can be normalized with respect to $q_{ab}$ (or to 
          $g_{ab}$) to form an orthonormal dyad.

\smallskip
\ni
This proposition implies that the Riemann surfaces $S_u$ are 
connected, simply connected subsets of a flat plane, and globally 
defined complex coordinates $(\zeta,\bar\zeta)$ can be introduced 
on one of the leaves, e.g. on $S_0$, which can be extended to the 
other leaves (e.g. by $X^aD_a\zeta=0$ for some vector field $X^a$ 
on $\Sigma$ which is transversal to the leaves) to obtain a 
coordinate system $(u,\zeta,\bar\zeta)$ on $\Sigma$. These 
coordinates are not unique: By appropriate coordinate transformation 
$M^a=(\partial/\partial\bar\zeta)^a$ can be achieved. Hence, in 
particular, $M^a$ and $\bar M^a$ are globally well defined on $S_u$ 
and their Lie bracket is zero. Thus $\zeta$ is the {\it 
anti}-holomorphic coordinate on the Riemann surfaces $S_u$. 
The complex coordinates $(\zeta,\bar\zeta)$ are still not unique: The 
remaining allowed transformation of them is $\zeta=\exp({\rm i}a(u))
\zeta'+A(u)$ for real $a(u)$ and complex $A(u)$. 
In the rest of the present paper we will assume that these 
convexity conditions are satisfied. 

\medskip
\ni
{$\bullet$ The geometry of the domain of dependence $D(\Sigma)$} 

\ni
Obviously the previous constructions can be repeated on the leaves
$\Sigma_t$ of a foliation of ${\rm int}\,D(\Sigma)$. If $\alpha$ is 
the lapse of this foliation, i.e. $\alpha t^a\nabla_at=1$, then the 
acceleration of $\Sigma_t$ is $a_b:=t^a\nabla_at_b=-D_b(\ln\alpha)$. 
Since $\Sigma$ is compact with boundary, each of the leaves $\Sigma_t$ 
of the foliation of $D(\Sigma)$ has the same 2-boundary: $\partial
\Sigma_t={\cal S}$. Thus, in particular, the lapse $\alpha$ is 
vanishing and the acceleration $a_e$ is diverging on ${\cal S}$. The 
general time derivative (of tensor fields) is defined as the 
projected Lie derivative along a vector field $\xi^a$ for which $\xi
^a\nabla_at=1$. Such a $\xi^a$ necessarily has the form $\xi^a=\alpha 
t^a+\beta^a$ (`evolution vector field'), where $\beta^a=P^a_b\beta^b$. 
Clearly, then the 1-parameter family of transformations generated by 
such a $\xi^a$ preserves the globally hyperbolic domain $D(\Sigma)$ 
precisely when $\beta^b$, the shift vector, is tangent to ${\cal S}$ 
on ${\cal S}$. The 2-surface ${\cal S}$ can be interpreted as the 
edge of the globally hyperbolic domain $D(\Sigma)$. 

Since $L^a$ is null it has the form $O^A\bar O^{A'}$ for some spinor 
field $O^A$, and if $L^a$ is constant then there exists a phase 
$\exp({\rm i}\psi)$ for which $\exp({\rm i}\psi)O^A$ is also constant. 
Thus we can (and will) assume that the spinor constituent $O^A$ of 
the constant null $L^a$ is constant. 

Using the single, fixed hypersurface $\Sigma$ and the constant null 
vector field $L^a$ one can introduce another coordinate system on 
${\rm int}\,D(\Sigma)$ in a natural way. Since the $u={\rm const}$ 
hypersurfaces are null, the integral curves of $L^a$ are geodesics. 
Let $r$ be the affine parameter along these geodesics measured from 
$\Sigma$. Then any $p\in{\rm int}\,D(\Sigma)$ sits on a uniquely 
determined integral curve of $L^a$ which intersects $\Sigma$ at a 
well defined point $q$. Then we label $p$ by the coordinates $(u,r,
\zeta,\bar\zeta)$ if the affine distance of $p$ from $q$ is $r$ and 
the coordinates of $q$ on $\Sigma$ are $(u,\zeta,\bar\zeta)$. 

Analogously, the Newman--Penrose complex null tetrad $\{L^a,N^a,
M^a,\bar M^a\}$ introduced on the single hypersurface $\Sigma$ can 
be extended to a tetrad field on the whole $D(\Sigma)$ by Lie 
propagating the vectors $N^a$ and $M^a$ along the integral curves 
of $L^a$. Since, however, $L^a$ is constant, this propagation 
implies that $L^e\nabla_eN^a=[L,N]^a=0$ and $L^e\nabla_eM^a=[L,M]
^a=0$, furthermore $[M,\bar M]^a=0$ still holds. Clearly, $L^a$, 
$M^a$ and $\bar M^a$ are coordinate vectors: $L^a=(\partial/\partial 
r)^a$, $M^a=(\partial/\partial\bar\zeta)^a$ and  $\bar M^a=(\partial
/\partial\zeta)^a$, but in general $N^a=(\partial/\partial u)^a-(H+
G\bar G)(\partial/\partial r)^a+\bar G(\partial/\partial\zeta)^a+G(
\partial/\partial\bar\zeta)^a$ for some real and complex functions 
$H=H(u,\zeta,\bar\zeta)$ and $G=G(u,\zeta,\bar\zeta)$, respectively. 
Then the metric of $D(\Sigma)$ in these coordinates takes the form 
$ds^2=2du\,dr-2d\zeta\,d\bar\zeta+2(G\,d\zeta+\bar G\,d\bar\zeta)\,
du+2H\,du^2$. By an appropriate allowed transformation of the 
coordinates $(\zeta,\bar\zeta)$ the function $G$ can be ensured to 
satisfy $(\partial G/\partial\bar\zeta)=(\partial\bar G/\partial
\zeta)$, whenever the constant spinor field $O^A$ can be completed 
by another spinor $I^A$ such that $O_AI^A=1$, $N^a=I^A\bar I^{A'}$ 
and $O^A\bar I^{A'}=M^a$ hold (see [11]). The remaining allowed 
transformations of the complex coordinates are $\zeta=\exp({\rm i}
a)\zeta'+A(u)$ for some real constant $a$ and complex function 
$A(u)$. The spinor field $I^A$ can be given explicitly as $I^A=
\Vert Z\Vert^{-1}t^{AA'}\bar O_{A'}$, and, conversely, $O^A=-2\Vert Z
\Vert t^{AA'}\bar I_{A'}$. This spin frame is uniquely determined 
and constant along the integral curves of $L^a$, i.e. $L^a\nabla_a
O_B=L^a\nabla_aI_B=0$. 

As an illustration consider the standard Cartesian coordinate system 
$(t,x,y,z)$ in Minkowski spacetime, let $\Sigma$ be the ball of radius 
$R$ in the $t=0$ hyperplane and ${\cal S}$ its boundary, a 2-sphere of 
radius $R$, which is clearly convex. $L_a:=\nabla_au$, $u:=t-z$, is 
constant and null, and the Riemann surfaces $S_u$ in $\Sigma$ are the 
$z={\rm const}$ 2-planes, in which $\zeta:={1\over\sqrt{2}}(x-{\rm 
i}y)$ are the complex anti-holomorphic coordinates. The boundary 
$\gamma_u$ of the Riemann surface $S_u$ is a circle in the 2-plane 
orthogonal to the $z$-axis with radius $\sqrt{R^2-u^2}$. The affine 
parameter along the integral curves of $L^a$, measured from $\Sigma$, 
is $r=t$. In these coordinates the spacetime metric takes the form 
$ds^2=2du\,dr-2d\zeta\,d\bar\zeta-du^2$.

\bigskip
\ni
{\lbf 3 Zero quasi-local mass configurations of matter fields}
\bigskip
\ni
{\bf 3.1 Zero quasi-local energy-momentum and zero quasi-local mass 
configurations}
\medskip
\ni
Let $\Sigma$ be any smooth, compact spacelike hypersurface with 
smooth boundary ${\cal S}:=\partial\Sigma$. Then for any vector field 
$K^a$ we can form the integral ${\tt Q}_\Sigma[K^a]:=\int_\Sigma K_aT
^{ab}t_b{\rm d}\Sigma$, where $t^a$ is the future pointing unit 
timelike normal to $\Sigma$ and ${\rm d}\Sigma:={1\over3!}t^e
\varepsilon_{eabc}$ is the induced volume element 3-form on $\Sigma$. 
(Here $\varepsilon_{abcd}$ is the spacetime volume 4-form.) In 
general ${\tt Q}_\Sigma[K^a]$ depends on $\Sigma$, but if $K^a$ is a 
Killing vector of the spacetime, then it depends only on the 
homology class of $\Sigma$ modulo its boundary ${\cal S}$: If 
$\Sigma'$ is another smooth compact spacelike hypersurface with 
the same boundary ${\cal S}$, then ${\tt Q}_\Sigma[K^a]$ and ${\tt Q}
_{\Sigma'}[K^a]$, defined on $\Sigma$ and $\Sigma'$, coincide. Hence 
the integral depends only on ${\cal S}$ and can be denoted by 
${\tt Q}_{\cal S}[K^a]$. In fact, ${\tt Q}_{\cal S}[K^a]$ can be 
rewritten as the 2-surface integral of some `superpotential 
2-form' ${1\over2}\cup[K^a]_{ab}$ on ${\cal S}$, where $K_eT^{ef}
\varepsilon_{fabc}=3\nabla_{[a}\cup[K^e]_{bc]}$. 

Thus suppose that $M$ is the Minkowski spacetime, fix a Cartesian 
coordinate system $\{x^{\ua}\}$, ${\ua}=0,...,3$, and write the 
general Killing 1-form as $K_a=T_{\ua}\nabla_ax^{\ua}+M_{\ua\ub}(x
^{\ua}\nabla_ax^{\ub}-x^{\ub}\nabla_ax^{\ua})$ for some constant 
coefficients $T_{\ua}$ and $M_{\ua\ub}=-M_{\ub\ua}$. Being ${\tt Q}
_{\cal S}[K^e]$ linear in $K_a$, it has the structure ${\tt Q}
_{\cal S}[K^e]=T_{\ua}{\tt P}^{\ua}+M_{\ua\ub}{\tt J}^{\ua\ub}$, 
and the coefficients of $T_{\ua}$ and $M_{\ua\ub}$ define the 
quasi-local energy-momentum ${\tt P}^{\ua}$ and angular momentum 
${\tt J}^{\ua\ub}$ of the matter fields, respectively, associated 
with the closed spacelike 2-surface ${\cal S}$. In fact, ${\tt P}
^{\ua}$ and ${\tt J}^{\ua\ub}$ are elements of the dual space of the 
space of the translation and boost-rotation Killing vectors of 
the Minkowski spacetime, respectively, and under the Poincar\'e 
transformations of the Cartesian coordinates, $x^{\ua}\mapsto x
^{\ub}\Lambda_{\ub}{}^{\ua}+C^{\ua}$, they transform just in the 
expected way. (For a more detailed discussion see [7].)  

Next suppose that the energy-momentum tensor satisfies the dominant 
energy condition: $T^{ab}V_b$ is future pointing and non-spacelike 
for any future pointing and non-spacelike vector $V^a$, and clarify 
the positivity properties of the energy-momentum. Then {\it 
${\tt P}^{\ua}$ is future pointing and non-spacelike} with respect to 
the natural Lorentz metric $\eta_{\ua\ub}:={\rm diag}(1,-1,-1,-1)$ on the 
space of translations, i.e. both the quasi-local mass ${\tt m}$, 
defined by ${\tt m}^2:={\tt P}^{\ua}{\tt P}^{\ub}\eta_{\ua\ub}$, and the 
quasi-local energy ${\tt P}^0$ are non-negative. If ${\tt P}^0=0$, then 
by the dominant energy condition the energy-momentum tensor must be 
vanishing on every spacelike hypersurface $\Sigma$ with the fixed 
2-boundary ${\cal S}$, implying the vanishing of the whole 
energy-momentum vector ${\tt P}^{\ua}$ as well. Thus {\it the 
quasi-local energy-momentum ${\tt P}^{\ua}$ is vanishing if and only 
if the energy-momentum tensor of the matter fields is vanishing 
on the whole domain of dependence $D(\Sigma)$ of $\Sigma$ with the 
given 2-boundary ${\cal S}$}. 

Similarly, let us consider the zero quasi-local mass configurations 
of the matter fields. If ${\tt m}=0$, then let $L_a:={\tt P}_{\ua}
\nabla_ax^{\ua}$, which is a constant null 1-form field on $M$; and 
if ${\tt P}^{0}$ is positive, then $L^a$ is future pointing. Then 

$$
0={\tt m}^2:={\tt P}^{\ua}{\tt P}^{\ub}\eta_{\ua\ub}=\int_\Sigma 
L_aT^{ab}t_b{\rm d}\Sigma. \eqno(3.1.1)
$$
Since by the dominant energy condition the integrand on the right 
is non-negative, this implies that $L_aT^{ab}=0$. 
Therefore, the energy-momentum tensor has a constant null eigenvector 
with zero eigenvalue, i.e. its algebraic type is pure 
radiation. Hence, again by the dominant energy condition, it has the 
form $T^{ab}=\tau L^aL^b$ for some non-negative function $\tau$ on 
$M$. This implies that ${\tt P}^{\ua}=eL^{\ua}$ and ${\tt J}^{\ua\ub}
=L^{\ua}e^{\ub}-L^{\ub}e^{\ua}$, where $e:=\int_\Sigma\tau L^at_a
{\rm d}\Sigma$ and $e^{\ua}:=\int_\Sigma\tau x^{\ua}L^at_a{\rm d}
\Sigma$. Conversely, if $T^{ab}$ is of pure radiation with constant 
null eigenvector $L^a$ with zero eigenvalue, then obviously ${\tt P}
^{\ua}$ is null, and hence ${\tt m}=0$. Therefore, {\it the 
quasi-local mass ${\tt m}$ is vanishing if and only if the 
energy-momentum tensor of the matter fields is of pure radiation 
with constant null eigenvector $L_a={\tt P}_{\ua}\nabla_ax^{\ua}$ on 
the whole domain of dependence $D(\Sigma)$}. 

In the next three subsections we discuss the consequences of these 
general results for specific matter fields. In particular, we 
determine the zero quasi-local mass configurations of the Higgs and 
Yang--Mills fields, which turn out to be special radiative solutions 
of the field equations. 

\bigskip
\ni
{\bf 3.2 Zero quasi-local mass configurations of the Higgs fields}
\medskip
\ni
Let $\phi^{\bi}$, ${\bi}=1,...,n$, be finite number of real scalar 
fields, $G_{\bi\bj}$ a symmetric constant $n\times n$ matrix, 
$V(\phi)$ an algebraic function of the fields, and consider the 
Lagrangian density $L_H:={1\over2}G_{\bi\bj}g^{ab}\nabla_a\phi^{\bi}
\nabla_b\phi^{\bj}-V(\phi)$. The multiplet $\phi^{\bi}$ of fields 
subject to the dynamics governed by this Lagrangian is called a Higgs 
field. (The fields $\phi^{\bi}$ can be considered as the components 
of a section in some frame field of a real vector bundle $H(M)$ of 
rank $n$ over $M$ associated with a principal $G$-bundle. But then 
they can be globally defined only if the frame field is globally 
defined, i.e. when $H(M)$ is globally trivializable. 
However, the section of $H(M)$ that $\{\phi^{\bi}\}$ represents can 
be globally defined and the whole analysis can be carried out even if 
$H(M)$ is not globally trivializable. In that case the small boldface 
Latin indices can be considered as abstract indices referring to the 
fiber space and $G_{\bi\bj}$ as a fiber metric, but its components can 
be chosen to be constant only on the domain of the trivializations 
of $H(M)$. In the presence of a gauge field $A^{\bi}_{a{\bk}}$, 
coupled minimally to the Higgs field, the spacetime covariant 
derivative $\nabla_a\phi^{\bi}$ is replaced by the spacetime {\it and} 
gauge-covariant derivative $\nabla_a\phi^{\bi}+eA^{\bi}_{a{\bk}}\phi
^{\bk}$ for some coupling constant $e$. The complex Higgs fields 
are considered as pairs of real ones. In particular, for a single 
{\it complex} scalar field $\varphi=\phi^{\bf 1}+{\rm i}\phi^{\bf 2}$ the 
metric $G_{\bi\bj}=\delta_{\bi\bj}$, ${\bi},{\bj}={\bf 1},{\bf 2}$, gives 
the familiar real scalar product $\varphi\bar\varphi$.) 
The corresponding field equations are $G_{\bj\bi}\nabla_a\nabla^a
\phi^{\bi}+2(\partial V/\partial\phi^{\bj})=0$. The symmetric 
energy-momentum tensor is $T_{ab}=G_{\bi\bj}(\nabla_a\phi^{\bi})
(\nabla_b\phi^{\bj})-{1\over2}g_{ab}g^{cd}G_{\bi\bj}(\nabla_c\phi
^{\bi})(\nabla_d\phi^{\bj})+g_{ab}V(\phi)$. For the sake of concreteness 
we assume that the `potential term' $V(\phi)$ has the form $V=
{1\over2}m^2G_{\bi\bj}\phi^{\bi}\phi^{\bj}+{1\over4}\lambda(G_{\bi\bj}
\phi^{\bi}\phi^{\bj})^2$, i.e. the fields are of rest-mass $m$ and if 
$\lambda\not=0$ then they are self-interacting. 

The energy-momentum tensor with this potential satisfies the dominant 
energy condition precisely when $\lambda\geq0$ and $G_{\bi\bj}$ is 
positive semi-definite, and then $T_{ab}=0$ is equivalent to $\phi
^{\bi}=0$ provided $G_{\bi\bj}$ is positive definite. (In contrast to 
quantum gauge theoretical investigations, in the present paper we 
assume that $m$ is real, and hence $m^2\geq0$. The positive 
definiteness of $G_{\bi\bj}$ is equivalent to that the actual 
representation of the gauge group $G$ is homomorphic to a subgroup of 
the {\it compact} $O(n)$.) Next suppose that $T_{ab}L^b=0$ for some 
null vector field $L^a$. Then, if $G_{\bi\bj}$ is positive definite, 
$T_{ab}L^aL^b=0$ implies that $L^a\nabla_a\phi^{\bi}=0$. Completing 
$L^a$ to a Newman--Penrose complex null tetrad $\{L^a,N^a,M^a,\bar 
M^a\}$ as e.g. in Section 2, this implies $\nabla_a\phi^{\bi}=f^{\bi}
L_a+g^{\bi}M_a+\bar g^{\bi}\bar M_a$ for some real valued field 
$f^{\bi}$ and complex $g^{\bi}$. Substituting $L^a\nabla_a\phi^{\bi}
=0$ back to $T_{ab}L^b=0$ we obtain that $2V(\phi)=G_{\bi\bj}\nabla
_a\phi^{\bi}\nabla^a\phi^{\bj}=-2G_{\bi\bj}g^{\bi}\bar g^{\bj}$. 
Since its left hand side is non-negative and the right hand side is 
non-positive, this implies that $g^{\bi}=0$, and if at least one of 
$m$ and $\lambda$ is non-zero then $\phi^{\bi}=0$. Therefore, {\it 
for positive definite $G_{\bi\bj}$ non-trivial, zero quasi-local mass 
configurations exist only for zero-rest-mass, non-self-interacting 
Higgs fields}. 
In this case $\nabla_a\phi^{\bi}=f^{\bi}L_a$ and $T_{ab}=G_{\bi\bj}f
^{\bi}f^{\bj}L_aL_b$, and {\it the zero quasi-local mass configurations 
are precisely those fields $\phi^{\bi}$ that are constant in the 
directions $L^a$, $M^a$ and $\bar M^a$}: $L^a\nabla_a\phi^{\bi}=0$ 
and $M^a\nabla_a\phi^{\bi}=\bar M^a\nabla_a\phi^{\bi}=0$. Thus the 
energy-momentum tensor of the zero quasi-local mass configurations 
of the Higgs field is the sum of the energy-momentum tensor of the 
decoupled, individual components of $\phi^{\bi}$ for each ${\bi}$. 
The directional derivatives of $f^{\bi}=N^a\nabla_a\phi^{\bi}$ are 
$L^a\nabla_af^{\bi}=0$ and $M^a\nabla_af^{\bi}=[M,N]^a\nabla_a\phi
^{\bi}+N^a\nabla_a(M^b\nabla_b\phi^{\bi})=[M,N]^a(L_aN^b+N_aL^b-M_a
\bar M^b-\bar M_aM^b)\nabla_b\phi^{\bi}=[M,N]^aL_af^{\bi}=0$. Thus 
in the coordinate system $(u,r,\zeta,\bar\zeta)$ of Section 2 this 
means that $\phi^{\bi}=\phi^{\bi}(u)$ and $f^{\bi}=f^{\bi}(u)$. In 
particular, if $\varphi$ is a zero-rest-mass {\it complex} scalar 
field with zero quasi-local mass, then $\nabla_a\varphi=fL_a$ for some 
{\it complex} $f$, and $L^a\nabla_a\varphi=M^a\nabla_a\varphi=\bar M^a
\nabla_a\varphi=0$ and $T_{ab}=f\bar fL_aL_b$ hold.

\bigskip
\ni
{\bf 3.3 Zero quasi-local mass configurations of the Yang--Mills 
fields}
\medskip
\ni
Let the Yang--Mills field be represented on a real vector bundle 
$F(M)$ of rank $N$, associated with a principal $G$-bundle, in a fixed 
frame field by the familiar connection 1-forms (or gauge potentials) 
$A^{\bK}_{a{\bL}}$ and the corresponding curvature 2-forms (or field 
strengths) $F^{\bK}{}_{{\bL}cd}:=\nabla_cA^{\bK}_{d{\bL}}-\nabla_d
A^{\bK}_{c{\bL}}+A^{\bK}_{c{\bM}}A^{\bM}_{d{\bL}}-A^{\bK}_{d{\bM}}
A^{\bM}_{c{\bL}}$. Thus the indices ${\bK},
{\bL},...=1,...,N$ are referring to a fixed (maybe only locally 
defined) frame field in $F(M)$, and $A^{\bK}_{a{\bL}}$ and $F^{\bK}
{}_{{\bL}ab}$ are 1 and 2-forms, respectively. They take their value 
in the matrix Lie algebra representing the Lie algebra ${\cal G}$ of 
the gauge group $G$, by means of which representation the vector 
bundle $F(M)$ was associated with the principal $G$-bundle. The 
dynamics of the Yang--Mills fields is governed by the Lagrangian 
$L_{YM}:=-{1\over4}F^{\bK}{}_{{\bL}cd}F^{\bL}{}_{\bK}{}^{cd}$. The 
corresponding field equations are $\nabla_bF^{\bK}{}_{\bL}{}^{ab}+
A^{\bK}_{b{\bM}}F^{\bM}{}_{\bL}{}^{ab}-F^{\bK}{}_{\bM}{}^{ab}A^{\bM}
_{b{\bL}}=0$, and the symmetric energy-momentum tensor is $T_{ab}=
-(F^{\bK}{}_{{\bL}ca}F^{\bL}{}_{\bK}{}^c{}_b-{1\over4}g_{ab}F^{\bK}
{}_{{\bL}cd}F^{\bL}{}_{\bK}{}^{cd})$. 
If $\{e^{\bK}_{\alpha{\bL}}\}$, $\alpha,\beta,...=1,...,\dim{\cal G}$, 
is a fixed basis of the Lie algebra in the given representation, 
then we can write $F^{\bK}{}_{{\bL}ab}=:e^{\bK}_{\alpha{\bL}}F^\alpha
{}_{ab}$ and $A^{\bK}_{a{\bL}}=:e^{\bK}_{\alpha{\bL}}A^\alpha_a$, and 
the real 2-forms $F^\alpha{}_{ab}$ can be represented by the 
symmetric spinors $\phi^\alpha_{AB}=\phi^\alpha_{(AB)}$ defined by 
$F^\alpha{}_{ab}=\varepsilon_{A'B'}\phi^\alpha_{AB}+\varepsilon_{AB}\bar
\phi^\alpha_{A'B'}$. (Of course, there is absolutely no connection 
between the spacetime spinor indices $A$, $B$, ... and the {\it 
boldface} bundle indices ${\bK}$, ${\bL}$, ... .) 
Defining the structure constants $c^\alpha_{\beta\gamma}$ of the Lie 
algebra of the gauge group by $e^{\bK}_{\beta{\bM}}e^{\bM}_{\gamma{\bL}}
-e^{\bK}_{\gamma{\bM}}e^{\bM}_{\beta{\bL}}=:c^\alpha_{\beta\gamma}e^{\bK}
_{\alpha{\bL}}$, the Bianchi identities and the Yang--Mills field 
equations can be rewritten as $\varepsilon^{CD}(\nabla_{CA'}\phi
^\alpha_{DA}+c^\alpha_{\beta\gamma}A^\beta_{CA'}\phi^\gamma_{DA})=0$. 
It might be interesting to note that the expression between the 
parentheses is just the spacetime and gauge covariant derivative of 
$\phi^\alpha_{AB}$, where the latter is defined on the vector bundle 
based on the adjoint representation of the gauge group. The 
energy-momentum tensor of the Yang--Mills fields takes the form 
$T_{ab}=2G_{\alpha\beta}\phi^\alpha_{AB}\bar\phi^\beta_{A'B'}$, where 
we introduced the notation $G_{\alpha\beta}:=e^{\bK}_{\alpha{\bL}}
e^{\bL}_{\beta{\bK}}=G_{(\alpha\beta)}$. In particular, if the 
representation of the gauge group on which the definition of the 
vector bundle $F(M)$ is based is the adjoint representation, then the 
frame field $e^{\bK}_{\alpha{\bL}}$ can be chosen to be the structure 
constants $c^\gamma_{\alpha\beta}$, and the metric $G_{\alpha\beta}$ 
will be the familiar Cartan--Killing metric on the Lie algebra ${\cal 
G}$. This metric is well known to be positive definite precisely for 
compact semisimple Lie groups. 

The energy-momentum tensor satisfies the dominant energy condition 
if $G_{\alpha\beta}$ is positive semi-definite, and in this case 
$T_{ab}=0$ is equivalent to the flatness of the connection $A^{\bK}
_{a{\bL}}$ provided $G_{\alpha\beta}$ is positive definite. Next 
suppose that $L^a=O^A\bar O^{A'}$ is an eigenvector of $T_{ab}$ with 
zero eigenvalue. If $G_{\alpha\beta}$ is positive definite then this 
implies that $\phi^\alpha_{AB}O^B=0$, and hence that $\phi^\alpha
_{AB}=\phi^\alpha O_AO_B$ for some complex functions $\phi^\alpha$. 
Then by the results of subsection 3.1, $O^A\bar O^{A'}$ is constant, 
and hence $O^A$ can be (and, in fact, will be) chosen to be 
constant. Thus, in particular, the algebraic type of the spinors 
$\phi^\alpha_{AB}$ is $N$ (see subsection 4.1 below) and $T_{ab}=
2G_{\alpha\beta}\phi^\alpha\bar\phi^\beta L_aL_b$. Substituting this 
form of $\phi^\alpha_{AB}$ into the field equations we find $O^A
(\nabla_{AA'}\phi^\alpha+c^\alpha_{\beta\gamma}A^\beta_{AA'}\phi
^\gamma)=0$, i.e. 

$$\eqalignno{
L^a\nabla_a\phi^\alpha&=c^\alpha_{\beta\gamma}\phi^\beta A^\gamma_a
 L^a, &(3.3.1a)\cr
M^a\nabla_a\phi^\alpha&=c^\alpha_{\beta\gamma}\phi^\beta A^\gamma_a
 M^a. &(3.3.1b)\cr}
$$
To solve these equations we should use the expression of the 
curvature $F^\alpha{}_{ab}$ in terms of the gauge potential 
$A^\alpha_a$. However, to simplify the calculations, it seems 
reasonable to use the maximal, allowed gauge fixing. 

Let $\Lambda^{\bK}{}_{\bL}$ be a gauge transformation in $F(M)$, i.e. 
a function on $M$ taking its value in the matrix Lie group 
representing the gauge group. Then under the action of this gauge 
transformation the gauge potential is well known to transform 
according to $A^{\bK}_{a{\bL}}\mapsto\bar A^{\bK}_{a{\bL}}:=\Lambda
_{\bM}{}^{\bK}A^{\bM}_{a{\bN}}\Lambda^{\bN}{}_{\bL}+\Lambda_{\bM}{}
^{\bK}(\nabla_a\Lambda^{\bM}{}_{\bL})$, where $\Lambda_{\bM}{}^{\bN}$ 
is defined by $\Lambda^{\bM}{}_{\bR}\Lambda_{\bM}{}^{\bN}=\delta
^{\bN}_{\bR}$. We want to use this freedom to make as many components 
of the gauge potential zero as it is possible. Thus suppose that 
$X^a$ and $Y^a$ are vector fields on $M$ whose contraction with the 
transformed gauge potential are vanishing, $X^a\bar A^{\bR}_{a{\bS}}
=Y^a\bar A^{\bR}_{a{\bS}}=0$, and clarify the conditions of the 
existence of such gauge transformations. Then by the transformation 
formula these conditions take the form 

$$
X^a\nabla_a\Lambda^{\bK}{}_{\bL}=-X^aA^{\bK}_{a{\bM}}\Lambda^{\bM}
{}_{\bL}, 
\hskip 20pt
Y^a\nabla_a\Lambda^{\bK}{}_{\bL}=-Y^aA^{\bK}_{a{\bM}}\Lambda^{\bM}
{}_{\bL}. \eqno(3.3.2)
$$
This is a system of partial differential equations for $\Lambda
^{\bK}{}_{\bL}$, and it is a direct calculation to check that its 
integrability condition is 

$$
F^{\bK}{}_{{\bL}ab}X^aY^b=-\Bigl(A^{\bK}_{a{\bM}}\Lambda^{\bM}{}
_{\bL}+\nabla_a\Lambda^{\bK}{}_{\bL}\Bigr)\bigl[X,Y\bigr]^a.
\eqno(3.3.3)
$$
Thus, if $X^a$ and $Y^a$ form an involutive distribution, this is 
equivalent to $F^{\bK}{}_{{\bL}ab}X^aY^b=0$. 

Next let us fix a spacelike hypersurface $\Sigma$ and a complex null 
tetrad $\{L^a,N^a,M^a,\bar M^a\}$ which is adapted to the foliation 
of $\Sigma$ by $S_u$ and Lie propagated along the integral curves of 
$L^a$, as in the last point of Section 2. Then the result that the 
anti-self-dual part of the field strength has the structure $\phi
^\alpha_{AB}=\phi^\alpha O_AO_B$ is equivalent to 

$$\eqalignno{
&F^{\bK}{}_{{\bL}ab}L^aM^b=0, \hskip 20pt F^{\bK}{}_{{\bL}ab}
  M^a\bar M^b=0,&(3.3.4a)\cr
&F^{\bK}{}_{{\bL}ab}L^aN^b=0, \hskip 20pt F^{\bK}{}_{{\bL}ab}
  N^a\bar M^b=-\phi^\alpha e^{\bK}_{\alpha{\bL}}.&(3.3.4b)\cr}
$$
Since for the vectors of the null tetrad $[L,M]^a=[M,\bar M]^a=0$ 
holds, by (3.3.3) and (3.3.4a) we can impose the gauge conditions 
$M^aA^{\bK}_{a{\bL}}=L^aA^{\bK}_{a{\bL}}=0$. Then (3.3.4b) reduces to 
$L^a\nabla_a(N^bA^{\bK}_{b{\bL}})=0$ and $\bar M^a\nabla_a(N^bA^{\bK}
_{b{\bL}})=-\phi^\alpha e^{\bK}_{\alpha{\bL}}$; while the field 
equations (3.3.1a) and (3.3.1b) reduce to the {\it linear} 
equations 

$$
L^a\nabla_a\phi^\alpha=0, \hskip 20pt M^a\nabla_a\phi^\alpha=0. 
\eqno(3.3.5)
$$
Thus the functions $\phi^\alpha$ are constant in the directions $L^a$ 
and $M^a$, i.e. in particular, they are anti-holomorphic on the 
Riemann surfaces $S_u$. Then, however, it is straightforward to write 
down the general solution of (3.3.4b)-(3.3.5) for the gauge potentials 
too. It is gauge equivalent to the gauge potential given by 

$$\eqalignno{
&L^aA^\alpha_a=M^aA^\alpha_a=0, &(3.3.6a)\cr
&N^aA^\alpha_a=\Phi^\alpha(u,\zeta)+\overline{\Phi^\alpha}(u,
\bar\zeta), &(3.3.6b)\cr}
$$
where $\Phi^\alpha(u,\zeta)$, $\alpha=1,...,\dim{\cal G}$, are 
arbitrary complex valued functions being smooth (in fact $C^2$ is 
enough) in $u$ and complex analytic in $\zeta$. Essentially this 
$\Phi^\alpha$ is a complex primitive function of the only independent 
component of the corresponding field strength: $\phi^\alpha(u,\zeta)
=-(\partial\Phi^\alpha(u,\zeta)/\partial\zeta)$. The gauge potential 
$A^\alpha_a$ satisfies the Lorenz gauge condition $\nabla^aA^\alpha_a
=0$, and the transformations preserving the gauge conditions (3.3.6a) 
are of the form $\Lambda^{\bK}{}_{\bL}=\Lambda^{\bK}{}_{\bL}(u)$. We 
call (3.3.6) a {\it pp}--wave solution of the Yang--Mills equations, 
though these solutions, discovered by Coleman [14], are known as 
`non-Abelian plane waves'. Indeed, these are genuine generalizations 
of the familiar plane wave solutions: For the latter the functions 
$\phi^\alpha$ are bounded in the whole Minkowski spacetime, and hence 
are independent of $\zeta$. However, quasi-locally $\phi^\alpha$ may 
be any anti-holomorphic function. Therefore, {\it the zero quasi-local 
mass configurations of the Yang--Mills equations with positive 
definite metric $G_{\alpha\beta}$ are the {\it pp}--waves (3.3.6), and, 
conversely, any {\it pp}-wave configuration has zero quasi-local 
mass}. The energy-momentum tensor for the {\it pp}-wave solutions is 
the sum of the energy-momentum tensors of the components of the 
gauge fields for each $\alpha$. 
The expression (3.3.6) provides an exact solution of the Yang--Mills 
field equations for arbitrary gauge group (i.e. even if the metric 
$G_{\alpha\beta}$ is {\it not} positive definite) even on a curved 
background spacetime admitting $L^a$ as a constant null vector field, 
but in these more general cases the solution 
(3.3.6) cannot be characterized as {\it the} zero quasi-local mass 
configurations: The quasi-local mass is zero for the {\it pp}--wave 
solutions, but there might be other solutions too giving zero 
quasi-local mass. The solution (3.3.6) is completely analogous to 
the {\it pp}-wave solutions of general relativity, which are known to 
be precisely the zero Dougan--Mason quasi-local mass configurations 
[10,11]. This result for the {\it non-Abelian} Yang--Mills fields 
gives further support to the claim [7] that any reasonable quasi-local 
energy-momentum expression in general relativity should be such that 
the zero-mass configurations be precisely the {\it pp}-waves. 

In the analogous representation of the electromagnetic field by 
the symmetric spinor $\phi_{AB}$, defined as the anti-self-dual 
part of the field strength by $F_{ab}=\varepsilon_{A'B'}\phi_{AB}+
\varepsilon_{AB}\bar\phi_{A'B'}$, the energy-momentum tensor is $T
_{ab}=2\phi_{AB}\bar\phi_{A'B'}$, while the source-free Maxwell 
equations take the form $\nabla_{A'}{}^A\phi_{AB}=0$. This $T_{ab}$ 
satisfies the dominant energy condition, and hence for the zero 
quasi-local mass configurations of the electromagnetic field one 
has $\phi_{AB}=\phi O_AO_B$ for some constant spinor field $O_A$. 
Then the Maxwell equations yield $O^A\nabla_{AA'}\phi=0$. Therefore, 
if $I^A$ is another (not necessarily constant) spinor field on 
$M$ for which $O_AI^A=1$ and $\{L^a,N^a,M^a,\bar M^a\}$ is the 
corresponding Newman--Penrose complex null tetrad, then {\it 
$\phi$ is constant in the directions $L^a$ and $M^a$}. 

The zero quasi-local mass configuration of the minimally coupled 
Yang--Mills--Higgs fields, formulated on the Whitney sum $H(M)
\oplus F(M)$ of the vector bundles above with the Lagrangian 
$L_H+L_{YM}$, can be treated analogously. These configurations are 
again {\it pp}-wave like solutions of the coupled Yang--Mills--Higgs 
field equations with a common (constant, null) wave vector $L^a$ of 
the Higgs and Yang--Mills fields. 

Finally, it could perhaps be worth noting that the overall picture 
remains the same for the coupled Einstein--Yang--Mills system. In 
fact, the general analysis of [9,10] can be completed by the 
analysis above to show that, under the conditions given there, the 
Dougan--Mason quasi-local mass is vanishing if and only if the 
geometry of $D(\Sigma)$ is given by the {\it pp}-wave solutions of 
the Einstein--Yang--Mills equations found by G\"uven [15]. 

\bigskip
\ni
{\bf 3.4 On the zero quasi-local mass configurations of spinor fields}
\medskip
\ni
The Lagrangian of the Weyl spinor field $\phi^A$ is $L_W={{\rm i}
\over2}(\bar\phi^{A'}\nabla_{AA'}\phi^A-\phi^A\nabla_{AA'}\bar\phi
^{A'})$, and the corresponding field equation and energy-momentum 
tensor, respectively, is $\nabla_{A'A}\phi^A=0$ and $T_{ab}={{\rm i}
\over4}(\bar\phi_{A'}\nabla_{BB'}\phi_A-\phi_A\nabla_{BB'}\bar\phi
_{A'}+\bar\phi_{B'}\nabla_{AA'}\phi_B-\phi_B\nabla_{AA'}\bar\phi
_{B'})$. Clearly, for the solutions of the form $\phi^A=\phi O^A$ 
for some {\it constant} spinor field $O^A$ the energy-momentum tensor 
has $L^a=O^A\bar O^{A'}$ as a null eigenvector with zero eigenvalue, 
and hence the corresponding quasi-local mass is zero. In this case 
$O^A\nabla_{AA'}\phi=0$, and hence $L^a\nabla_a\phi=0$ and $M^a
\nabla_a\phi=0$, i.e. $\phi=\phi(u,\zeta)$. 
However, since this energy-momentum tensor does {\it not} satisfy 
even the weak energy condition, the vanishing of the quasi-local 
mass for the Weyl neutrino field does not imply $T_{ab}L^a=0$ for 
some null $L^a$, and that the zero quasi-local mass configurations 
would necessarily be some wave--type solutions. 

Similarly, the energy-momentum tensor for the zero-rest-mass (or 
simply z.r.m.) spinor fields with index structure $\phi_{A_1...A
_{m+1}}^{B'_1...B'_m}=\phi_{(A_1...A_{m+1})}^{(B'_1...B'_m)}$, $m=
0,1,2,...$ and $2s=2m+1$, does not satisfy the week energy condition 
(see [16]). Non-zero rest mass for the spinor fields does not 
improve the positivity properties of the energy-momentum tensor: 
For the Dirac spinor $\Psi^\alpha$, represented by the pair $(\phi^A,
\bar\psi^{A'})$ of Weyl spinors, the energy-momentum tensor is the 
sum of the energy-momentum tensors for the constituent spinors 
$\phi^A$ and $\bar\psi^{A'}$ above, independently of the rest-mass 
parameter in the Lagrangian for $\Psi^\alpha$. 

Since the energy-momentum tensor for other higher (i.e. $s\geq
{3\over2}$) spin fields (e.g. with the index structure $\phi_{A_1
...A_{2s}}=\phi_{(A_1...A_{2s})}$, including the Weyl spinor $\phi
_{ABCD}$ for the linearized gravity [12]) cannot even be defined, 
their special `pure radiative configurations' does not seem to be 
connected with a special value of some quasi-local observable, e.g. 
with the vanishing of the quasi-local mass. Nevertheless, as we will 
see in the next section, their pure radiative modes can in fact be 
characterized quasi-locally just in an analogous way. 

\bigskip

\ni
{\lbf 4 Linear zero-rest-mass fields with any spin}
\bigskip
\ni
{\bf 4.1 Generalities}
\medskip
\ni
We learnt in the previous section that the vanishing of the 
quasi-local mass is equivalent to special radiative solutions of 
the z.r.m. scalar and Yang--Mills field equations, whenever the 
field equations became {\it linear}. As we will see, these are 
special {\it linear zero-rest-mass} field equations. A particularly 
interesting special case of the linear z.r.m. field equations is the 
field equations for the gravitational perturbations of the Minkowski 
spacetime. These motivate us to consider these equations in general 
rather than only e.g. Weyl spinors, the classical representatives of 
fermionic matter. 

If $\phi_{A_1...A_{2s}}$, $2s=0,1,2,...$, is a completely symmetric 
spinor field, then let $\alpha_A$, $\beta_A$, ..., $\zeta_A$ be the 
principal spinors of $\phi_{A_1...A_{2s}}$, i.e. for which $\phi_{A_1
...A_{2s}}=\alpha_{(A_1}\beta_{A_2}...\zeta_{A_{2s})}$ holds (see 
e.g. [12,17]). If all these principal spinors coincide, then 
$\phi_{A_1...A_{2s}}$ is called null or of type N, and hence it has 
the structure $\phi_{A_1...A_{2s}}=\alpha_{A_1}...\alpha_{A_{2s}}$. 
The linear z.r.m. field equation for any $2s\in{\bf N}$ is $\nabla
_{A'}{}^A\phi_{AA_2...A_{2s}}=0$, which in Minkowski spacetime implies 
the wave equation $\nabla_{BB'}\nabla^{BB'}\phi_{A_1...A_{2s}}=0$. By 
linear z.r.m. field equation for $s=0$ one usually means the wave 
equation, the field equation that we got in subsection 3.2. for the 
Higgs field in the special case $m=\lambda=0$. 

If $\Sigma$ is a spacelike hypersurface with normal $t^a$, then the 
unitary spinor form of the differential operator $P^b_a\nabla_b$, 
acting on the unprimed spinor fields defined on $\Sigma$, is well 
known [18] to be ${\cal D}_{AB}:=\sqrt{2}t^{A'}{}_{(A}\nabla_{B)A'}$. 
Then the 3+1 form of the linear z.r.m. field equations for non-zero 
spin takes the form 

$$\eqalignno{
{1\over\sqrt{2}}t^e\nabla_e\phi_{A_1...A_{2s}}+{\cal D}_{B(A_1}
  \phi^B{}_{A_2...A_{2s})}&=0, &(4.1.1)\cr
{\cal D}_{BC}\phi^{BC}{}_{A_3...A_{2s}}&=0. &(4.1.2)\cr}
$$
The first expresses the time derivative of the spinor field in 
terms of its spatial derivative, while the second is a restriction 
on the spinor field on the spacelike hypersurface. Thus, foliating 
$D(\Sigma)$ by smooth spacelike Cauchy surfaces $\Sigma_t$ with normal 
$t^a$ and lapse $\alpha$, and multiplying both sides of (4.1.1) by 
$\alpha$, we find that it is the evolution equation in $D(\Sigma)$ 
along the special `evolution vector field' $\alpha t^a$; and (4.1.2) 
is a constraint. Hence the Weyl spinor field ($s={1\over2}$) is 
unconstrained, but for $s\geq1$ equation (4.1.2) gives $2s-1$ 
restrictions on the $2s+1$ components of the spinor field, yielding 
two complex unconstrained degrees of freedom at each point of 
$\Sigma$. In particular, for the Maxwell field the familiar electric 
and magnetic field strengths are given by $E_a+{\rm i}B_a=2\varepsilon
_{A'B'}\phi_{AB}t^{BB'}$, thus by $D_a(E^a+{\rm i}B^a)=2t^{BB'}(\nabla
_{B'}{}^A\phi_{AB})=\sqrt{2}{\cal D}_{AB}\phi^{AB}$ the constraint 
(4.1.2) is nothing but the conditions that the field strengths be 
divergence free, and hence the unconstrained degrees of freedom are 
the two components of such $E^a$s and the two components of such 
$B^a$s. (For a {\it complex} scalar field $\phi$ satisfying the wave 
equation the freely specifiable initial data on $\Sigma$ consists of 
$\phi$ and $t^a\nabla_a\phi$, thus we have two complex degrees of 
freedom at each point of $\Sigma$ for $s=0$ too.) 

Let $C^\infty(\Sigma,{\bf S}_{(A_1...A_{2s})})$ denote the infinite 
dimensional complex vector space of the smooth totally symmetric 
spinor fields with indices $2s$ on $\Sigma$, and, for any $\phi_{A_1
...A_{2s}}$, $\psi_{A_1...A_{2s}}\in C^\infty(\Sigma,{\bf S}_{(A_1...
A_{2s})})$, define the Hermitian scalar product 

$$
\langle\phi_{A_1...A_{2s}},\psi_{A_1...A_{2s}}\rangle_s:=\int_\Sigma
\phi_{A_1...A_{2s}}\bar\psi_{A'_1...A'_{2s}}2^st^{A_1A'_1}...
t^{A_{2s}A'_{2s}}{\rm d}\Sigma. \eqno(4.1.3)
$$
Clearly, this is positive definite and let $\Gamma_s$ denote the 
metric completion of $C^\infty(\Sigma,{\bf S}_{(A_1...A_{2s})})$ in 
the corresponding $L_2$-norm $\Vert\phi_{A_1...A_{2s}}\Vert^2:=\langle 
\phi_{A_1...A_{2s}},\phi_{A_1...A_{2s}}\rangle_s$. Thus $\Gamma_s$ is a 
Hilbert space. The differential operator ${\cal D}:C^\infty(\Sigma,
{\bf S}_{(A_1...A_{2s})})\rightarrow C^\infty(\Sigma,{\bf S}_{(A_1...A
_{2s-2})})$ $:\phi_{A_1...A_{2s}}\mapsto {\cal D}^{BC}\phi_{BCA_1...A
_{2s-2}}$ is a linear, underdetermined elliptic operator, and the 
closure of the space of the solutions of (4.1.2), $\hat\Gamma_s:=
\overline{\ker{\cal D}}$, can be interpreted as the `constraint 
surface' in $\Gamma_s$. It might be interesting to note that the 
formal adjoint of ${\cal D}$ is just the complex conjugate 3-surface 
twistor operator. 

\bigskip
\ni
{\bf 4.2 pp-wave and plane wave solutions}
\medskip
\ni
The significance of type $N$ solutions is that they are thought of 
as purely radiative configurations. In the present section, however, 
we consider only a special subclass of null fields, especially since 
the zero quasi-local mass configurations of the Yang--Mills fields 
belong to this. 

Thus suppose that $\phi_{A_1...A_{2s}}$ is not only null, but also a 
{\it pp}-wave as well in the sense that its $2s$-fold principal 
spinor is proportional to a {\it constant} spinor $O^A$, i.e. $\phi
_{A_1...A_{2s}}=\phi O_{A_1}...O_{A_{2s}}$. 
(For the Yang--Mills field this structure of the anti-self-dual 
curvature $\phi^\alpha_{AB}$ for some {\it constant} $O_A$ also was a 
consequence of ${\tt m}=0$.) Note that the {\it pp}-wave solutions 
can exist precisely when the spacetime geometry is also a {\it 
pp}-wave, i.e. admits a constant null vector field, e.g. in 
Minkowski spacetime. 
(Indeed, if $(M,g_{ab})$ admits a constant spinor field $O^A$, then 
$L^a:=O^A\bar O^{A'}$ is a constant null vector field, and, 
conversely, if $O{}^A\bar O{}^{A'}$ is constant, then there exists a 
phase $\exp({\rm i}\psi)$ for which $\exp({\rm i}\psi)O^A$ is constant. 
If $(M,g_{ab})$ admits {\it two independent} {\it pp}-waves [in the 
sense that the two constant spinors are not proportional to each 
other], then $(M,g_{ab})$ is flat.) 
Then the field equations yield that $O^A\nabla_{AA'}\phi=0$, which, 
in the Newman--Penrose complex null frame considered in Section 2, 
gives $L^a\nabla_a\phi=0$ and $M^aD_a\phi=0$. These are precisely the 
field equations (3.3.5) for the quasi-local mass configurations of 
the Yang--Mills fields in a special gauge. 
Therefore, in the coordinates introduced at the end of Section 2, we 
have that $\phi$ is independent of the coordinates $r$ and $\bar\zeta$ 
(i.e. in particular it is anti-holomorphic on the Riemann surfaces 
$S_u$), and hence $\phi=\phi(u,\zeta)$. Substituting $\phi O_{A_1}...
O_{A_{2s}}$ into (4.1.2) we find that, for {\it pp}-waves, (4.1.2) is 
equivalent to $M^a\nabla_a\phi=0$, but (4.1.1) is to both $M^a\nabla
_a\phi=0$ and $L^a\nabla_a\phi=0$. In particular, for the Maxwell 
field (and for the components of the Yang--Mills fields for each 
value of the index $\alpha$) $E^a+{\rm i}B^a=-2\phi\Vert Z\Vert M^a$, 
implying that $E_aE^a=B_aB^a$, $E_aB^a=0$, $(E_a+{\rm i}B_a)Z^a=0$, 
the well known characteristic properties of the null Maxwell fields, 
and that $E_aE^a+B_aB^a=-4\vert\phi\vert^2\Vert Z\Vert^2$. 

As we saw in subsection 3.2, in the zero quasi-local mass 
configurations of the (real or complex) massless {\it scalar} fields 
the field variable $\phi$ was not only anti-holomorphic, but constant 
on the $u={\rm const}$ `wave fronts'. Thus it is natural to 
consider this special case too. We call the {\it pp}-wave solution 
$\phi O_{A_1}...O_{A_{2s}}$ a plane wave if $\phi$ is constant on the 
`wave-fronts', i.e. when $\phi=\phi(u)$. The familiar Fourier modes 
in Minkowski spacetime, indexed by the future pointing constant 
null vector field $L^a$, have the form $\phi_{A_1...A_{2s}}(L,x):=\Phi
_{A_1...A_{2s}}(L)\exp(-{\rm i}L_{\ua}x^{\ua})$. Here $\Phi_{A_1...A_{2s}}
(L)$ is a constant spinor depending on $L^a$, $x^{\ua}$ are the 
Cartesian coordinates and $L_{\ua}$ are the components of $L_a$ in 
the Cartesian basis: $L_a=L_{\ua}\nabla_ax^{\ua}$. This is a solution 
of the linear z.r.m. field equations precisely when $L^{A'A}\Phi_{
AA_2...A_{2s}}(L)=0$, and, if we write $L^{AA'}=O^A\bar O^{A'}$, then 
$\Phi_{A_1...A_{2s}}(L)=c(L^e)\,O_{A_1}...O_{A_{2s}}$ for some constant 
$c(L^e)$. Furthermore, in the coordinates $(u,r,\zeta,\bar\zeta)$ 
adapted to a {\it constant} complex null tetrad $\{L^a,N^a,M^a,\bar 
M^a\}$ one has $\exp(-{\rm i}L_{\ua}x^{\ua})=\exp(-{\rm i}u)$, i.e. the 
Fourier mode $\phi_{A_1...A_{2s}}(L,x)$ is a special plane wave.

\bigskip
\ni
{\bf 4.3 pp-wave spinor fields and quasi-local observables}
\medskip
\ni
Although we do not have any energy-momentum tensor for a general 
spinor field, on a given spacelike hypersurface $\Sigma$ we can 
define the `strength' of the fields by the integral of their 
pointwise positive definite Hermitian norm that we used to define 
the $L_2$ scalar product (4.1.3) above. This norm motivates us to 
introduce, for totally symmetric unprimed spinor fields, the tensor 
field $T_{a_1...a_{2s}}$ $:=2^s\phi_{A_1...A_{2s}}\bar\phi_{A'_1...
A'_{2s}}$, $2s\in{\bf N}$. This is totally symmetric and trace free, 
and if $\phi_{A_1...A_{2s}}$ is a solution of the linear z.r.m. field 
equation then it is divergence-free. For $s=1$ it coincides with the 
{\it physical} energy-momentum tensor for the Maxwell field above, 
and for $s=2$ with the Bel--Robinson (or, according to the convention 
of [12], four times the Bel--Robinson) tensor for the linearized 
Weyl tensor of the weak gravitational field. Furthermore, it satisfies 
an analog of the dominant energy condition: The proof of the analogous 
statement for the energy-momentum tensor of the Maxwell field given 
in [12] can be applied directly to $T_{a_1...a_{2s}}$. Indeed, since 
any future pointing nonspacelike vector is a sum of two future 
pointing null vectors, it is enough to prove that $T^a{}_{a_2...a_{2s}}
X^{a_2}_2...X^{a_{2s}}_{2s}$ is future pointing and non-spacelike only 
for future pointing {\it null} vectors $X^a_2$, ..., $X^a_{2s}$. Thus 
let $X^a_i=\xi^A_i\bar\xi^{A'}_i$, $i=2,...,2s$. Then $T^a{}_{a_2...a
_{2s}}X^{a_2}_2...X^{a_{2s}}_{2s}=2^s(\phi^A{}_{A_2...A_{2s}}\xi^{A_2}
_2...\xi^{A_{2s}}_{2s})(\bar\phi^{A'}{}_{A'_2...A'_{2s}}\bar\xi^{A'_2}
_2...\bar\xi^{A'_{2s}}_{2s})$, which is future pointing and null. 
Hence for any vector field $K^a$ it is natural to define the 
quasi-local norm of the totally symmetric spinor fields {\it with 
respect to $K^a$} on $\Sigma$ as the flux integral of the current 
$K^at^{a_2}...t^{a_{2s-1}}T_{aa_2...a_{2s-1}}{}^e$ on $\Sigma$: 

$$
{\tt N}_\Sigma\bigl[K^a\bigr]:=\int_\Sigma K^aT_{aa_2...a_{2s}}t^{a_2}
...t^{a_{2s}}{\rm d}\Sigma. \eqno(4.3.1)
$$
Clearly, for $K^a=t^a$ this reduces to the $L_2$ norm $\Vert\phi
_{A_1...A_{2s}}\Vert^2$. In contrast to ${\tt Q}_{\cal S}[K^a]$ of 
subsection 3.1, in general it depends not only on $\partial\Sigma$ 
but on $\Sigma$ as well (i.e. not `conserved'), even for a Killing 
vector $K^a$. Obviously, ${\tt N}_\Sigma[K^a]$ is non-negative for any 
future pointing non-spacelike $K^a$. 

In spite of the fact that ${\tt N}_\Sigma[K^a]$ is not conserved, 
it can be used to characterize the identically vanishing and the 
{\it pp}-wave spinor fields on the globally hyperbolic domain 
$D(\Sigma)$: 

\medskip
\ni
{\bf Theorem 4.3.2:}  

\ni
The linear z.r.m. field $\phi_{A_1...A_{2s}}$ is vanishing on 
$D(\Sigma)$ if and only if ${\tt N}_{\Sigma_0}[K^a]=0$ for some (and 
hence for every) future pointing {\it timelike} $K^a$ on some (and 
hence on every) Cauchy surface $\Sigma_0$ for $D(\Sigma)$. In 
particular, $\phi_{A_1...A_{2s}}=0$ on $D(\Sigma)$ iff $\Vert\phi_{A_1
...A_{2s}}\Vert=0$ on $\Sigma$. 

\smallskip
\ni
{\it Proof:}
\item{} {If ${\tt N}_{\Sigma_0}[K^a]=0$ for some future pointing {\it 
  timelike} $K^a$, then, since $T^a{}_{a_2...a_{2s}}t^{a_2}...t^{a_{2s}}$ 
  is future pointing and non-spacelike, by (4.3.1) $T_{aa_2...a_{2s}}
  t^{a_2}...t^{a_{2s}}=0$ must hold. But then, contracting with $t^a$, 
  we obtain that $\phi_{A_1...A_{2s}}=0$ on $\Sigma_0$. However, 
  foliating the domain of dependence $D(\Sigma)$ of $\Sigma$ by a 
  family $\Sigma_t$ of Cauchy surfaces such that e.g. $\Sigma=\Sigma
  _0$, by (4.1.1) this implies that $\phi_{A_1...A_{2s}}=0$ on the 
  whole $D(\Sigma)$. To see this it is enough to recall that any 
  $C^2$ solution $\phi_{A_1...A_{2s}}$ of the linear z.r.m. field 
  equation solves the wave equation, for which a well posed initial 
  value problem can be formulated. However, its solution is causal 
  in the sense that the vanishing of the field and its first time 
  derivative on the initial hypersurface $\Sigma$ implies the 
  vanishing of the field on the whole $D(\Sigma)$. But by (4.1.1) 
  the vanishing of $\phi_{A_1...A_{2s}}$ on $\Sigma$ implies the 
  vanishing of $t^e\nabla_e\phi_{A_1...A_{2s}}$ as well, and hence 
  $\phi_{A_1...A_{2s}}=0$ on $D(\Sigma)$. (For a more detailed 
  discussion of the initial value problem for the wave equation 
  see e.g. [19].) The proof in the other direction is trivial. \sq}

\bigskip

\ni
{\bf Theorem 4.3.3:}  

\ni
The linear z.r.m. field $\phi_{A_1...A_{2s}}$ is a {\it pp}-wave on 
$D(\Sigma)$ if and only if ${\tt N}_{\Sigma_0}[L^a]=0$ for some future 
pointing constant null $L^a$ on some (and hence on any) Cauchy 
surface $\Sigma_0$ in $D(\Sigma)$. 

\smallskip
\ni
{\it Proof:}
\item{}{Suppose that ${\tt N}_{\Sigma_0}[L^a]=0$ for some future 
 pointing non-spacelike $L^a$ on $\Sigma_0$, but $\phi_{A_1...A_{2s}}$ 
 is not identically zero on $D(\Sigma)$. Then, since both $L^a$ and 
 $T^a{}_{a_2...a_{2s}}t^{a_2}...t^{a_{2s}}$ are future pointing and 
 non-spacelike, by (4.3.1) they must be null and proportional to 
 each other. Thus let $L^a=O^A\bar O^{A'}$. Then 

$$
0=L^aT_{aa_2...a_{2s}}t^{a_2}...t^{a_{2s}}=2^s\bigl(O^A\phi_{A
A_2...A_{2s}}\bigr)\bigl(\bar O^{A'}\bar\phi_{A'A'_2...A'_{2s}}
\bigr)t^{A_2A'_2}...t^{A_{2s}A'_{2s}}.\eqno(4.3.4)
$$
 This implies that $O^A\phi_{AA_2...A_{2s}}=0$ on $\Sigma_0$. Hence 
 $\phi_{A_1...A_{2s}}$ is null and $O^A$ is (proportional to) its 
 $2s$-fold principal spinor on $\Sigma_0$. Thus it has the form 
 $\phi _{A_1...A_{2s}}=\phi O_{A_1}...O_{A_{2s}}$ for some complex 
 function $\phi$. Since by our assumption $O^A$ is constant on 
 $\Sigma_0$, $\phi_{A_1...A_{2s}}$ is a {\it pp}-wave on $\Sigma_0$. 
 Next we show that it is a {\it pp}-wave on the whole $D(\Sigma)$ as 
 well. Let us extend $O^A$ from $\Sigma_0$ to $D(\Sigma)$ as a constant 
 spinor field, denoted also by $O^A$, and let the corresponding 
 null vector field on $D(\Sigma)$ be denoted also by $L^a$. Obviously, 
 this extension is unique. Finally, let us extend the complex 
 function $\phi$ from $\Sigma_0$ to $D(\Sigma)$ to be constant along 
 the integral curves of $L^a$, i.e. by $L^a\nabla_a\phi=0$, and define 
 $\phi_{A_1...A_{2s}}=\phi O_{A_1}...O_{2s}$. This solves the linear 
 z.r.m. field equations on $D(\Sigma)$ if $M^a\nabla_a\phi=0$ as 
 well. Thus we should show that this solution is unique, and hence 
 that $\phi_{A_1...A_{2s}}$ on $D(\Sigma)$ is a {\it pp}-wave. Since, 
 however, every solution of the linear z.r.m. field equations is a 
 solution of the wave equation on $D(\Sigma)$ as well, and the 
 solution of the initial value problem for the wave equation on a 
 globally hyperbolic domain is unique, the uniqueness of the solution 
 $\phi_{A_1...A_{2s}}=\phi O_{A_1}...O_{2s}$ follows.} 

\item{}{\hskip 20pt Conversely, if $\phi_{A_1...A_{2s}}$ is null with 
 $2s$-fold principal spinor $\alpha^A$, in particular if it is a 
 {\it pp}-wave, then obviously ${\tt N}_{\Sigma_0}[\alpha^A\bar\alpha
 ^{A'}]=0$ on any Cauchy surface $\Sigma_0$ for $D(\Sigma)$. \sq}

\smallskip 
\ni
If ${\tt N}_\Sigma[K^a]=0$ for two independent future pointing and 
nonspacelike vector fields $K^a_1$ and $K^a_2$ on $\Sigma$, then 
$K^a_1+K^a_2$ is timelike and ${\tt N}_\Sigma[K^a_1+K^a_2]=0$, and 
hence by Theorem 4.3.2 $\phi_{A_1...A_{2s}}=0$ on $D(\Sigma)$. In 
Minkowski spacetime the space of the constant spacetime vector 
fields on the spacelike hypersurface $\Sigma$ is isomorphic to the 
Lorentzian vector space of the translation Killing fields: 
The restriction to $\Sigma$ of a translation Killing field is 
constant on $\Sigma$, and, conversely, every such constant vector 
field is the restriction to $\Sigma$ of some uniquely determined 
translation. Therefore, restricting the vector field $K^a$ in the 
argument of ${\tt N}_\Sigma[K^a]$ to be a {\it constant} vector field 
on $\Sigma$, $K_a$ has the form $T_{\ua}\nabla_ax^{\ua}$ for some 
constants $T_{\ua}$ and the Cartesian coordinates $x^{\ua}$ in $M$. 
Consequently, (4.3.1) takes the form ${\tt N}_\Sigma[K^a]={\tt N}
_\Sigma^{\ua}T_{\ua}$. The coefficient ${\tt N}_\Sigma^{\ua}$, being 
an element of the dual of the space of the constant spacetime vector 
fields on $\Sigma$, can be interpreted as some form of the quasi-local 
energy-momentum, and we call it the `norm-energy-momentum'. Its 
length with respect to the Lorentzian metric on the space of the 
constant vector fields, $\eta_{\ua\ub}{\tt N}_\Sigma^{\ua}{\tt N}
_\Sigma^{\ub}$, might be called the `norm-mass'. Indeed, by theorems 
4.3.2 and 4.3.3 the role that they play in the characterization of 
the special z.r.m. configurations on $D(\Sigma)$ is analogous to that 
of the physical quasi-local energy-momentum and mass. In particular, 
for {\it pp}-wave configurations with non-zero spin, characterized by 
$(O^A,\phi)$ on $\Sigma$, the quasi-local norm-energy-momentum takes 
the form 

$$
{\tt N}^{\ua}_\Sigma=L^{\ua}2^s\int_\Sigma\vert\phi\vert^2\Vert Z
\Vert^{2s-1}{\rm d}\Sigma, \eqno(4.3.5)
$$
where $L^{\ua}$ are the components of $L^a=O^A\bar O^{A'}$ in the 
Cartesian coordinate system. Interestingly enough, ${\tt Q}_{\cal S}
[K^a]$, built from the {\it physical} energy-momentum tensor of the 
scalar field, can also be recovered from (4.3.5) for $s=0$ provided 
the $\phi\mapsto f\Vert Z\Vert$ substitution is made, where the 
function $f$ measures the gradient of $\phi$: It is given by $\nabla
_a\phi=fL_a$. 

On the other hand, in general ${\tt N}_\Sigma^{\ua}$ depends on 
$\Sigma$ (i.e. `non-conserved') and its physical dimension is 
different from that of energy-momentum. In particular, for the 
gravitational perturbations described by a spin-two field $\phi
_{ABCD}$, the spinor form of the linearized Weyl tensor of the 
weak gravitational field, the `norm-energy-density' $T_{abcd}t^at^b
t^ct^d$ has dimension $cm^{-4}$, and hence there are no powers $A$ 
and $B$ for which $G^Ac^BT_{abcd}t^at^bt^ct^d$ would have the correct 
dimension $g\,cm^{-1}sec^{-2}$ of the energy density in the 
traditional units. (Here $c$ is the speed of light in vacuum.)  

Clearly, the norm (4.1.3) and the tensor field $T_{a_1...a_{2s}}$ 
can be defined for any spinor field $\phi_{A_1...A_kB'_1...B'_l}$ 
for which $k+l=2s$. Then, however, $T_{a_1...a_{2s}}$ does not share 
the desirable properties that we have for totally symmetric purely 
primed or unprimed spinor fields. 
Another, potentially interesting generalization ${\tt N}_\Sigma
[K^a_1,...,K^a_k]$ of the norm may be introduced using a given 
collection $K^a_1$, ..., $K^a_k$, $k\leq(2s-1)$, of vector fields 
as the integral of $K^{a_1}_1...K^{a_k}_kT_{a_1...a_kb_{k+1}...
b_{2s}}t^{b_{k+1}}...t^{t_{2s}}$. A remarkable property of this 
expression for $k=2s-1$, $s\geq1$, is that it is conserved if 
$K^a_1$, ..., $K^a_k$ are conformal Killing fields. 

\bigskip

\ni
{\lbf 5 Two-surface characterization of classical fields}
\bigskip
\ni
{\bf 5.1 The holographic data for the pp-waves}
\medskip
\ni
Since the function $\phi=\phi(u,\zeta)$ in the {\it pp}-wave solutions 
of the linear z.r.m. field equations with any spin was complex 
analytic in the variable $\zeta$, any {\it pp}-wave solution on 
$\Sigma$, and hence on the whole domain of dependence $D(\Sigma)$ too, 
is completely determined by the constant spinor field $O^A$ and the 
value $\varphi$ of $\phi$ on ${\cal S}$, i.e. by the pair $(O^A,
\varphi)$. The aim of the present subsection is to clarify whether or 
not such a pair can be characterized in terms of structures and 
objects defined {\it only} on the 2-surface ${\cal S}$, and to 
determine the independent unconstrained 2-surface data on ${\cal S}$ 
corresponding to a {\it pp}-wave on $D(\Sigma)$. We call such a data 
set the {\it holographic data} for the {\it pp}-wave. 

Let us discuss first the spinor field $O^A$. 
The restriction to ${\cal S}$ of the constant spinor fields of the 
Minkowski spacetime can be recovered as the {\it constant spinor 
fields on ${\cal S}$}, where $\lambda_A$ is called constant on ${\cal 
S}$ if $\Pi^b_c\nabla_b\lambda_A=0$. In fact, in a general spacetime 
there are no constant spinor fields on ${\cal S}$, but in Minkowski 
spacetime the space of the constant spinor fields is two complex 
dimensional and inherits a natural symplectic metric. Thus they 
form an $SL(2,{\bf C})$ spin space $({\bf S}^A_0,\varepsilon_{AB})$. 
Since the restriction to ${\cal S}$ of the two linearly independent 
constant spinor fields of the Minkowski spacetime are constant on 
${\cal S}$ too, the solutions of $\Pi^b_c\nabla_b\lambda_A=0$ are 
precisely these restrictions. 
Thus if $O^A$ is constant on ${\cal S}$, then $O^A\bar O^{A'}$ 
determines a unique constant null {\it spacetime} vector field (the 
`null wave vector'), and hence a foliation of ${\cal S}$ by the 
level sets of the function $u:{\cal S}\rightarrow{\bf R}$ for which 
$\Pi^b_aO_B\bar O_{B'}=\delta_au$. 
By the convexity of ${\cal S}$ the vector field $z^a=\Pi^a_bO^B\bar 
O^{B'}$ is vanishing precisely at two points, $p_-$ and $p_+$. Thus 
the function $u$ is strictly monotonically increasing along its 
integral curves from $u_-$ to $u_+$. 
(Since $z^a$ is vanishing at $p_\pm$, actually these integral curves 
never reach the points $p_\pm$. Using the coordinate freedom $\zeta
=\zeta'+A(u)$, where $A(u)$ is an arbitrary complex valued function, 
the origin of the complex coordinates $(\zeta,\bar\zeta)$ in $S_u$ can 
be shifted to a point of the boundary $\partial S_u={\cal S}\cap S_u$. 
In particular, for appropriately chosen $A(u)$ the one-parameter 
family of the origins of the coordinates $(\zeta,\bar\zeta)$, i.e. in 
the coordinate system $(u,\zeta,\bar\zeta)$ the curve $u\mapsto(u,0,0)$, 
can be chosen to be an integral curve of $z^a$.) 
Again by the convexity of ${\cal S}$ the boundaries $\partial S_u=
{\cal S}\cap S_u$ are closed smooth curves, denoted by $\gamma_u$, 
which are the integral curves of the vector field $s^a=\varepsilon
^{ab}O_B\bar O_{B'}$. However, instead of the natural arch length 
parameter $s$ of the integral curves $\gamma_u$ it seems more 
convenient to use the parameter $w\in[0,2\pi)$ defined by $w\,{\rm 
Length}(\gamma_u):=2\pi s$, measured in each $S_u$ from the 
intersection point of $S_u$ and an integral curve of $z^a$, e.g. 
from the origin of the complex coordinates $(\zeta,\bar\zeta)$. Thus 
$(u,w)$ is a coordinate system on ${\cal S}-\{p_-,p_+\}\approx(u_-,u_+)
\times S^1$, and the coordinate lines $u={\rm const}$, i.e. the curves 
$\gamma_u$, can be specified in the complex coordinates as $S^1
\rightarrow S_u:w\mapsto(z_u(w),\bar z_u(w))$. 

Next consider the restriction of the functions $\phi$ to ${\cal S}$. 
If $\varphi:{\cal S}\rightarrow{\bf C}$ is the restriction to ${\cal 
S}$ of the type $N$ solution $\phi$ of the linear z.r.m. equation 
with the constant wave co-vector $O_A\bar O_{A'}=\nabla_au$, then 
this is not an arbitrary, general smooth complex valued function 
on ${\cal S}$. Indeed, on $\gamma_u={\cal S}\cap S_u$ it is the 
restriction from $S_u$ to $\gamma_u$ of a function which is {\it 
anti-holomorphic} on $S_u$. Hence we should characterize these 
special complex valued functions on ${\cal S}$. We show that the 
real and the imaginary parts of such a $\varphi$ are not quite 
independent. Thus let $F:={\rm Re}\,\phi$, the real part of $\phi$. 
This is harmonic w.r.t. the flat metric Laplace operator on the 
Riemann surfaces $S_u$ (and hence, in particular, $F$ is real 
analytic). Therefore, it is completely determined by its restriction 
$f:=F\vert_{\cal S}$ to ${\cal S}$ via the Dirichlet problem on each 
of the leaves $S_u$. Using the fact that each $S_u$ is connected and 
simply connected, standard theorems of complex analysis guarantee 
the existence of a real valued function $G$ on $\Sigma$ such that $G$ 
is harmonic and $F+{\rm i}\,G$ is anti-holomorphic on each $S_u$. Thus, 
in particular, $G$ is real analytic on the Riemann surfaces $S_u$. 
This $G$ is uniquely determined by $F$ up to an additive constant on 
each $S_u$, i.e. $G$ is unique up to addition of a function of $u$ 
only. Thus the free, unconstrained part of the boundary value 
$\varphi$ is e.g. the real part $f={\rm Re}\,\varphi={\rm Re}\,\phi
\vert_{\cal S}$ and the value of its imaginary part $g:={\rm Im}\,
\varphi={\rm Im}\,\phi\vert_{\cal S}$ on the integral curve of $z^a$ 
representing the origin of the complex coordinates. 
Therefore, in the coordinates $(u,w)$ above, the unconstrained part 
of $\varphi$ is equivalent to the pair $(f(u,w),g(u,0))$ of real 
functions. (Another, obviously equivalent, representation of the 
same $\varphi$ is given by the imaginary part of $\varphi$ and the 
value of its real part on the `origin curve'.) 
Finally, by Cauchy's integral formula the field $\phi$ at the point 
$(u,r,\zeta,\bar\zeta)$ of ${\rm int}\,D(\Sigma)$ can be recovered as 

$$
\phi(u,r,\zeta,\bar\zeta)=\phi(u,\zeta)={1\over2\pi{\rm i}}\int^{2\pi}_0
{f(u,w)+{\rm i}g(u,w)\over z_u(w)-\zeta} \dot z_u(w){\rm d}w, 
\eqno(5.1.1)
$$
where the complex contour is $\gamma_u:S^1\rightarrow S_u:w\mapsto
(z_u(w),\bar z_u(w))$, as above, and the dot denotes derivative with 
respect to the parameter $w$. For plane waves $f$ is independent of 
$w$, and hence the right hand side of (5.1.1) reduces to $f(u)+{\rm 
i}g(u)$. For linear z.r.m. fields with spin zero, for which the {\it 
pp}-waves are always plane waves, this is always the case. 

If $\Sigma$ is flat then the linear z.r.m. field equations take a 
simple, coupled system of linear equations for the various components 
of the spinor field. Penrose gave both algebraically general and 
special solutions to them in terms of complex contour integrals of 
appropriately chosen meromorphic functions [20], which contour 
integrals could be reformulated in the projective twistor space 
[21,17,22]. These contour integrals are similar to (5.1.1), but the 
contour here is in the spacetime. For special meromorphic functions 
the solutions of the linear z.r.m. field equations are called 
elementary states [21,17,22]. These are special solutions that are 
singular on the light cone of a point of the (compactified) 
Minkowski spacetime. The significance of these elementary states is 
that they form a dense subset in the space of all fields [17,23]. 
This raises the question as whether the {\it pp}-wave solutions 
play similar role in the space of all fields. 

\bigskip

\ni
{\bf 5.2 The density of the pp-wave modes in $\Gamma_s$}
\medskip
\ni
Because of the rather special nature of the {\it pp}-waves the 
significance of the previous investigations and the results may 
appear at first sight to be quite limited. If $\Sigma$ were a complete 
flat Euclidean 3-space, then the solutions $\phi_{A_1...A_{2s}}$ of 
the linear z.r.m. field equations that have finite $L_2$--norm (and 
hence bounded on the whole $\Sigma$) could be expanded as a Fourier 
integral of plane wave modes. The restriction of $\phi_{A_1...A_{2s}}$ 
to a compact piece of $\Sigma$ would, of course, be a solution there. 
However, quasi-locally there are much more solutions of the field 
equations: For example, the {\it pp}-wave solutions which are {\it 
not} plane waves are not bounded on an infinite $\Sigma$. 
Similarly, if $\Sigma$ were closed (i.e. compact with no boundary, 
e.g. a flat torus $\Sigma\approx S^1\times S^1\times S^1$), then any 
solution on $\Sigma$ would be periodic in the spatial directions, 
and could be expanded as a Fourier series of plane wave solutions. 
Thus, again, the space of solutions in the closed case is definitely 
`smaller' than in the quasi-local case. On the other hand, as we 
mentioned at the end of the previous subsection, the elementary 
states do form a dense subset in the space of all fields, and, in 
particular, in the space of solutions of the linear z.r.m. field 
equations. Moreover, quasi-locally there are much more `elementary' 
solutions, the {\it pp}-waves, that could be used as a basis to 
expand the elements of the much larger solution space. Therefore, 
the question arises naturally as whether or not the plane wave (or 
at least the {\it pp}-wave) solutions span the space of solutions 
in the quasi-local case too. The main goal of the present 
subsection is to show that every spinor field on $\Sigma$ can be 
uniformly approximated with arbitrary accuracy in an appropriate 
way by {\it pp}-wave solutions. The precise statement is the 
following theorem. 

\medskip
\ni
{\bf Theorem 5.2.1:}  

\ni
Let $O^A$ be a fixed constant spinor field. Then for any continuous 
totally symmetric spinor field $\psi_{A_1...A_{2s}}$, $s>0$, on 
$\Sigma$ and $\varepsilon>0$ there exist finitely many {\it pp}-wave 
solutions $\phi^iO_A$, $\chi^iO_A$, $i=1,...,N$, of the Weyl equation 
and totally symmetric constant spinor fields $C_i{}_{A_1...A_{2s}}$ 
such that 

$$
\vert\psi_{A_1...A_{2s}}(p)-\sum_{i=1}^N\phi^i(p)\bar\chi^i(p)
C_i{}_{A_1...A_{2s}}\vert<\varepsilon
$$
at each point $p\in\Sigma$. Here $\vert\, .\,\vert$ is the pointwise 
Hermitian norm defined by $\vert\psi_{A_1...A_{2s}}\vert^2:=2^s\psi
_{A_1...A_{2s}}\bar\psi_{A'_1...A'_{2s}}$ $t^{A_1A'_1}...t^{A_{2s}A'
_{2s}}$. 

\smallskip
\ni
{\it Proof:}
\item{}{ 
Let us fix a globally defined spin frame $\{{\cal E}^A_{\uA}\}$, 
${\uA}=0,1$, on $\Sigma$, and denote the components of any spinor 
field $\psi_{A_1...A_{2s}}$ in this basis by $\psi_{{\uA}_1...{\uA}
_{2s}}$, or simply by $\psi$. Thus $\psi$ is a typical component of 
the spinor field, which is a continuous complex valued function 
for continuous $\psi_{A_1...A_{2s}}$. The logic of our proof follows 
that of the proof of the Stone--Weierstrass theorem for real functions 
given in [24] and certain points of the proof for complex functions 
given in [25]. (We quote the Stone--Weierstrass theorem in the real 
case in the Appendix from [24].)}

\item{}{\hskip 20pt 
Let $(u,\zeta,\bar\zeta)$ be the globally defined coordinate system 
on $\Sigma$ determined by $O^A$, and let $\Phi(O^A)$ denote the set 
of complex valued functions $\phi=\phi(u,\zeta)$ that are smooth in 
their $u$ variable and complex analytic in their $\zeta$ variable. 
If $c_{m_1...m_k}\in{\bf C}$, $m_j=0,...,M_j$, $j=1,...,k$ and 
$$
P\bigl(z^1,...,z^k\bigr):=\sum_{m_1=0}^{M_1}...\sum_{m_k=0}^{M_k}
c_{m_1...m_k}\bigl(z^1\bigr)^{m_1}...\bigl(z^k\bigr)^{m_k}, 
$$
a $k$ complex variable polynomial, then for any $\phi^1,...,\phi^k
\in\Phi(O^A)$ the function $P(u,\zeta):=P(\phi^1(u,\zeta),...,$ $\phi
^k(u,\zeta))$ is smooth in $u$ and complex analytic in $\zeta$. Thus 
$\Phi(O^A)$ is an algebra with respect to the usual vector space 
operations and the pointwise multiplication, and clearly it is 
unital. Furthermore, $\Phi(O^A)$ separates points of $\Sigma$ in the 
sense that for any two different points $p,q\in\Sigma$ there exists 
a function $\phi$ in $\Phi(O^A)$ such that $\phi(p)\not=\phi(q)$. In 
fact, if $p$ and $q$ are in the same Riemann surface $S_u$, then e.g. 
$\phi=\zeta$, while if $u(p)\not=u(q)$ then any strictly monotonic 
real function $\phi=\phi(u)$ separates them.}

\item{}{\hskip 20pt 
Let $\Pi(O^A)$ denote the set of the complex valued functions of the 
form $\pi(u,\zeta,\bar\zeta):=\phi^1(u,\zeta)\bar\chi^1(u,\bar\zeta)+
...+\phi^n(u,\zeta)\bar\chi^n(u,\bar\zeta)$ on $\Sigma$, where $\phi^i$, 
$\chi^i\in\Phi(O^A)$. Obviously, the functions $\pi$ are continuous. 
Clearly, $\Pi(O^A)$ is the unital *--algebra generated by $\Phi(O^A)$, 
where the *--operation is the complex conjugation: The pointwise 
multiplication and the linear combination of any two functions 
$\pi_1$ and $\pi_2$ from $\Pi(O^A)$ by complex numbers belong to $\Pi
(O^A)$, and if $\pi\in\Pi(O^A)$ then its complex conjugate $\bar\pi$ 
also belongs to $\Pi(O^A)$. $\Phi(O^A)$ is a complex subalgebra in 
$\Pi(O^A)$.}

\item{}{\hskip 20pt 
Finally, let $\overline{\Pi}(O^A)$ denote the uniform closure of 
$\Pi(O^A)$, i.e. let $\overline{\Pi}(O^A)$ be the set of those 
functions $\rho:\Sigma\rightarrow{\bf C}$ such that for any 
$\varepsilon>0$ there exists a function $\pi\in\Pi(O^A)$ for which 
$\vert\rho(p)-\pi(p)\vert<\varepsilon$ $\forall p\in\Sigma$. (For 
functions or complex numbers $\vert\,.\,\vert$ means, of course, 
absolute value.) Our claim is to show that $\overline{\Pi}(O^A)$ is 
the space $C^0(\Sigma,{\bf C})$ of the continuous complex functions 
on $\Sigma$. To prove this, we need the following lemma.}

\medskip
\ni
{\bf Lemma 5.2.2:} 
\ni
\item{i.}   Any function $\rho\in\overline{\Pi}(O^A)$ is continuous; 
\item{ii.}  $\overline{\Pi}(O^A)$ is closed with respect to the uniform 
               closure; 
\item{iii.} $\overline{\Pi}(O^A)$ is a unital *--algebra. 

\smallskip
\ni
{\it Proof of Lemma 5.2.2:}
\item{i.}{ 
If $\rho\in\overline{\Pi}(O^A)$, then for any $\varepsilon>0$ $\exists
\pi\in\Pi(O^A)$ such that $\vert\rho(p)-\pi(p)\vert<\varepsilon$ for any 
$p\in\Sigma$. Since $\pi$ is continuous, $p$ has an open neighbourhood 
$U_p\subset\Sigma$ such that $\vert\pi(p)-\pi(q)\vert<\varepsilon$ for 
any $q\in U_p$. Then, however, $\vert\rho(p)-\rho(q)\vert=\vert\rho(p)-
\pi(p)+\pi(p)-\pi(q)+\pi(q)-\rho(q)\vert\leq\vert\rho(p)-\pi(p)\vert+
\vert\pi(p)-\pi(q)\vert+\vert\pi(q)-\rho(q)\vert<3\varepsilon$, i.e. 
$\rho$ is continuous. }

\item{ii.}{ 
If $\psi:\Sigma\rightarrow{\bf C}$ is a function such that for any 
$\varepsilon>0$ there exists a function $\rho\in\overline{\Pi}(O^A)$ 
satisfying $\vert\psi(p)-\rho(p)\vert<\varepsilon$ $\forall$ 
$p\in\Sigma$, then by the definition of $\overline{\Pi}(O^A)$ there 
exists a function $\pi\in\Pi(O^A)$ satisfying $\vert\pi(p)-\rho(p)
\vert<\varepsilon$ $\forall$ $p\in\Sigma$. Therefore, $\vert\psi(p)-
\pi(p)\vert=\vert\psi(p)-\rho(p)+\rho(p)-\pi(p)\vert\leq\vert\psi(p)-
\rho(p)\vert+\vert\rho(p)-\pi(p)\vert<2\varepsilon$, i.e. 
$\psi\in\overline{\Pi}(O^A)$. }

\item{iii.}{ 
Let $\rho_1$, $\rho_2\in\overline{\Pi}(O^A)$ and $\pi_1$, $\pi_2\in\Pi
(O^A)$ the corresponding functions approximating them better than 
$\varepsilon>0$. 
Then $\vert\bar\rho_1(p)-\bar\pi_1(p)\vert=\vert\rho_1(p)-\pi_1(p)\vert
<\varepsilon$ $\forall$ $p\in\Sigma$, i.e. $\overline{\Pi}(O^A)$ is 
closed with respect to complex conjugation: $\rho_1\in\overline
{\Pi}(O^A)$ implies $\bar\rho_1\in\overline{\Pi}(O^A)$. 
If $c_1,c_2\in{\bf C}$, then $\vert(c_1\rho_1+c_2\rho_2)(p)-(c_1\pi_1+
c_2\pi_2)(p)\vert\leq(\vert c_1\vert+\vert c_2\vert)\varepsilon$, i.e. 
$\overline{\Pi}(O^A)$ is a complex vector space. 
Finally, $\vert(\rho_1\rho_2)(p)-(\pi_1\pi_2)(p)\vert=\vert\rho_1(p)(
\rho_2(p)-\pi_2(p))+(\rho_1(p)-\pi_1(p))\pi_2(p)\vert\leq(\sup_{p\in
\Sigma}\vert\rho_1(p)\vert+\sup_{p\in\Sigma}\vert\pi_2(p)\vert)
\varepsilon$, i.e. $\overline{\Pi}(O^A)$ is an algebra. Since the 
constant functions belong to $\Phi(O^A)\subset\Pi(O^A)\subset\overline
{\Pi}(O^A)$, $\overline{\Pi}(O^A)$ is unital. \sq}

\smallskip
\ni
{\it Continuation of the proof of Theorem 5.2.1:}
\item{}{
By the first statement of Lemma 5.2.2 $\overline{\Pi}(O^A)\subset 
C^0(\Sigma,{\bf C})$, and by the third statement $\overline{\Pi}(O^A)$ 
has the structure $\overline{\Pi}(O^A)=\overline{\Pi}_{re}(O^A)+{\rm i}
\overline{\Pi}_{re}(O^A)$, where $\overline{\Pi}_{re}(O^A):=\{\rho\in
\overline{\Pi}(O^A)\vert\,\bar\rho=\rho\,\}$, the set of the real 
elements of $\overline{\Pi}(O^A)$. 
$\overline{\Pi}_{re}(O^A)$ is a real, unital subalgebra, it is 
closed with respect to the uniform closure, and by the first 
statement of Lemma 5.2.2 $\overline{\Pi}_{re}(O^A)\subset C^0(\Sigma,
{\bf R})$. We show that $C^0(\Sigma,{\bf R})\subset\overline{\Pi}_{re}
(O^A)$ holds as well, and hence that $C^0(\Sigma,{\bf C})=\overline
{\Pi}(O^A)$. The rest of the proof is essentially that given in 
[24], but for the sake of completeness we briefly repeat it here. 
This is based on the following two lemmas. }

\medskip
\ni
{\bf Lemma 5.2.3:}
\ni
\item{i.}  For any two different points $p$ and $q$ of $\Sigma$ and 
          real numbers $\alpha$ and $\beta$ there exists a function 
          $\rho\in\overline{\Pi}_{re}(O^A)$ such that $\rho(p)=\alpha$ 
          and $\rho(q)=\beta$; 
\item{ii.} If $\rho_1,...,\rho_k\in\overline{\Pi}_{re}(O^A)$, then the 
          upper and lower enveloping functions of $\{\rho_1,...,\rho_k
          \}$, defined pointwise by $\rho_1\cup...\cup\rho_k(p):=\max\{
          \rho_1(p),...,\rho_k(p)\}$ and $\rho_1\cap...\cap\rho_k(p):=
          \min\{\rho_1(p),...,\rho_k(p)\}$, respectively, are 
          elements of $\overline{\Pi}_{re}(O^A)$. 

\smallskip
\ni
{\it Proof of Lemma 5.2.3:}
\item{i.}{ 
Since $\Phi(O^A)$ separates points of $\Sigma$, there exists a 
function $\phi\in\Phi(O^A)\subset\overline{\Pi}(O^A)$ such that $\phi(p)
\not=\phi(q)$, and hence ${\rm Re}\,\phi(p)\not={\rm Re}\,\phi(q)$ or 
${\rm Im}\,\phi(p)\not={\rm Im}\,\phi(q)$. Since $\overline{\Pi}(O^A)$ 
is a *--algebra and ${\rm Re}\,\phi={1\over2}(\phi+\bar\phi)$ and ${\rm 
Im}\,\phi={1\over2{\rm i}}(\phi-\bar\phi)$, both ${\rm Re}\,\phi$ and 
${\rm Im}\,\phi$ belong to $\overline{\Pi}_{re}(O^A)$. Let $\sigma$ denote 
either ${\rm Re}\,\phi$ or ${\rm Im}\,\phi$ for which $\sigma(p)\not=
\sigma(q)$. Since $\overline{\Pi}_{re}(O^A)$ is unital, the function 
$\rho(s):=((\sigma(s)-\sigma(q))\alpha+(\sigma(p)-\sigma(s))\beta)/(
\sigma(p)-\sigma(q))$, $s\in\Sigma$, belongs to $\overline{\Pi}_{re}
(O^A)$ and has the desired properties. }

\item{ii.}{ 
It is enough to show the statement for $k=2$, whenever $\rho_1
\cup\rho_2={1\over2}(\rho_1+\rho_2+\vert\rho_1-\rho_2\vert)$ and $\rho_1
\cap\rho_2={1\over2}(\rho_1+\rho_2-\vert\rho_1-\rho_2\vert)$. Thus it is 
enough to show that $\vert\rho\vert\in\overline{\Pi}_{re}(O^A)$ for 
any $\rho\in\overline{\Pi}_{re}(O^A)$. For, let $M:=\sup_{p\in\Sigma}
\vert\rho(p)\vert$, which is finite. Then by the classical 
Weierstrass theorem (see Appendix) the real function $y\mapsto
\vert y\vert$ on $[-M,M]$ can be approximated uniformly by a 
polynomial $P(y)$ with accuracy $\varepsilon$, and hence with 
$y=\rho(p)$ one has $\vert P\circ\rho(p)-\vert\rho(p)\vert\,\vert=\vert 
P(y)-\vert y\vert\,\vert<\varepsilon$. Thus $\vert\rho\vert$ can be 
approximated uniformly on $\Sigma$ by the polynomial $P\circ\rho$ 
of $\rho$, which belongs to $\overline{\Pi}_{re}(O^A)$. Therefore, 
$\vert\rho\vert\in\overline{\Pi}_{re}(O^A)$. 
\sq}

\medskip
\ni
{\bf Lemma 5.2.4:}  

\ni
Let $f:\Sigma\rightarrow{\bf R}$ be any continuous function and ${\cal 
F}$ a set of continuous functions $F:\Sigma\rightarrow{\bf R}$ such 
that for any $p\in\Sigma$ there is a function $F\in{\cal F}$ 
satisfying $F(p)>f(p)$. Then there are finitely many functions 
$F_1$,...,$F_k\in{\cal F}$ such that $F_1\cup...\cup F_k(p)>f(p)$ 
for all $p\in\Sigma$. Similarly, if ${\cal G}$ is a set of functions 
such that for any $p\in\Sigma$ $\exists\,G\in{\cal G}$ such that 
$G(p)<f(p)$, then $\exists\,G_1,...,G_l\in{\cal G}$ such that $G_1\cap
...\cap G_l(p)<f(p)$ for all $p\in\Sigma$. 
\smallskip
\ni
{\it Proof of Lemma 5.2.4:}
\item{}{ 
Let $p\in\Sigma$ and $F_p\in{\cal F}$ such that $F_p(p)>f(p)$. Since 
both $f$ and $F_p$ are continuous, $p$ has an open neighbourhood 
$U_p\subset\Sigma$ such that $F_p(q)>f(q)$ for all $q\in U_p$. But 
these neighbourhoods form an open covering $\{U_p\vert\,p\in\Sigma\,
\}$ of the compact $\Sigma$, thus there are finitely many points $p_1,
...,p_k$ such that $\{U_{p_1},...,U_{p_k}\}$ already covers $\Sigma$. 
Then, however, $F_{p_i}(q)>f(q)$ $\forall q\in U_{p_i}$, $i=1,...,k$, 
implies $F_{p_1}\cup...\cup F_{p_k}(q)>f(q)$ for all $q\in\Sigma$. The 
proof of the second part of this lemma is similar. 
\sq}

\smallskip
\ni
{\it Continuation of the proof of Theorem 5.2.1:}
\item{}{
Returning to the proof of $C^0(\Sigma,{\bf R})\subset\overline{\Pi}_{re}
(O^A)$, let $f:\Sigma\rightarrow{\bf R}$ be any continuous function 
and $\varepsilon>0$. Then by the first statement of Lemma 5.2.3 for 
any two points $p,q\in\Sigma$ there exists a function $\rho_{pq}
\in\overline{\Pi}_{re}(O^A)$ such that 
$$\eqalignno{
&\rho_{pq}\bigl(p\bigr)>f\bigl(p\bigr)-\varepsilon, &(a)\cr
&\rho_{pq}\bigl(q\bigr)<f\bigl(q\bigr)+\varepsilon. &(b)\cr}
$$
Now fix $q$ and consider ${\cal F}_q:=\{\rho_{pq}\vert\,p\in\Sigma
\,\}$. Then the function $f-\varepsilon$ and ${\cal F}_q$ satisfy 
the conditions of Lemma 5.2.4, and hence for some finitely many 
points $p_1$,...,$p_k$ and the corresponding functions $\rho_{p
_1q},...,\rho_{p_kq}\in{\cal F}_q$ one has 
$$
\rho_q\bigl(p\bigr):=\rho_{p_1q}\cup...\cup\rho_{p_kq}\bigl(p\bigr)>
f\bigl(p\bigr)-\varepsilon 
\hskip 20pt \forall p\in\Sigma. \eqno(c)
$$
Then by the second statement of Lemma 5.2.3 $\rho_q\in\overline{\Pi}
_{re}(O^A)$, and by inequality (b) 
$$
\rho_q\bigl(q\bigr)=\max\{\rho_{p_1q}\bigl(q\bigr),...,\rho_{p_kq}
\bigl(q\bigr)\}<f\bigl(q\bigr)+\varepsilon. \eqno(d)
$$
Let ${\cal G}:=\{\rho_q\vert\,q\in\Sigma\,\}$. Then the function $f+
\varepsilon$ and ${\cal G}$ satisfy the conditions of Lemma 5.2.4 
for the lower enveloping case, i.e. for any $p\in\Sigma$ there 
exists a function $\rho_p\in{\cal G}$ such $\rho_p(p)<f(p)+
\varepsilon$, and hence there exist functions $\rho_{p_1},...,\rho
_{p_l}\in{\cal G}$ such that 
$$
\rho\bigl(p\bigr):=\rho_{p_1}\cap...\cap\rho_{p_l}\bigl(p\bigr)<
f\bigl(p\bigr)+\varepsilon 
\hskip 20pt \forall p\in\Sigma. 
$$
By the second statement of Lemma 5.2.3 $\rho\in\overline{\Pi}_{re}
(O^A)$ and, by inequality (c), $\rho$ satisfies 
$$
\rho\bigl(p\bigr)=\min\{\rho_{p_1}\bigl(p\bigr),...,\rho_{p_l}
\bigl(p\bigr)\}>f\bigl(p\bigr)-\varepsilon 
\hskip 20pt \forall p\in\Sigma.
$$
Thus, to summarize the last two inequalities, we obtain $\vert 
f(p)-\rho(p)\vert<\varepsilon$ for all $p\in\Sigma$, and hence $f\in
\overline{\Pi}_{re}(O^A)$. Therefore, $C^0(\Sigma,{\bf C})=\overline{\Pi}
(O^A)$. }

\item{}{\hskip 20pt
Thus we have proven that for any $\psi\in C^0(\Sigma,{\bf C})$ and 
any $\varepsilon>0$ there exist functions $\phi^1,...,\phi^n,\chi^1,
...,$ $\chi^n\in\Phi(O^A)$ such that $\vert\psi(p)-(\phi^1(p)\bar\chi^1
(p)+...+\phi^n(p)\bar\chi^n(p))\vert<\varepsilon$ for all $p\in\Sigma$, 
i.e. each component $\psi_{{\uA}_1...{\uA}_{2s}}$ of a continuous 
spinor field $\psi_{A_1...A_{2s}}$ can be approximated uniformly 
on $\Sigma$ by a function of the form $\sum_{i=1}^n\phi^i\bar\chi^i$ 
with arbitrary accuracy. But $\psi_{A_1...A_{2s}}$ has only finite 
number of components, thus there is a collection of functions 
$\phi^i,\chi^i\in\Phi(O^A)$, $i=1,...,N$, and constants $C_i{}_{{\uA}_1
...{\uA}_{2s}}$ (with value 0 or 1 depending on the indices $i$ 
and ${\uA}_1,...,{\uA}_{2s}$) such that for any set of spinor name 
indices ${\uA}_1,...,{\uA}_{2s}$ 
$$
\vert\psi_{{\uA}_1...{\uA}_{2s}}\bigl(p\bigr)-\sum_{i=1}^N\phi^i
\bigl(p\bigr)\bar\chi^i\bigl(p\bigr)C_i{}_{{\uA}_1...{\uA}_{2s}}\vert
<\varepsilon \hskip 20pt \forall p\in\Sigma. \eqno(e)
$$
Let the spinor basis $\{{\cal E}^{\uA}_A\}$ be chosen to be constant 
on $\Sigma$, whenever $C_i{}_{A_1...A_{2s}}:=C_i{}_{{\uA}_1...{\uA}_{2s}}
{\cal E}^{{\uA}_1}_{A_1}...{\cal E}^{{\uA}_{2s}}_{A_{2s}}$ are constant 
spinor fields. Then inequality (e) is almost the statement of 
Theorem 5.2.1 to be proven. The only difference between them is 
that $\vert\,.\,\vert$ has different meaning in Theorem 5.2.1 and in 
inequality (e): In the former it is the pointwise Hermitian norm 
defined with respect to the normal $t^a$ of $\Sigma$, while in (e) 
it is the absolute value of the components of the spinor fields in 
a given spinor basis $\{{\cal E}^{\uA}_A\}$. However, since the two 
norms are equivalent at each point $p\in\Sigma$ and $\Sigma$ is 
compact, the two norms are equivalent on $\Sigma$ too. This 
completes the proof of Theorem 5.2.1.  \sq}

\smallskip
\ni
Though at first sight this theorem is simply the Stone--Weierstrass 
theorem applied to spinor fields, it is slightly more than this. It 
tells us that the {\it pp}-wave solutions e.g. of the Weyl equation, 
even with a given single common constant wave spinor $O^A$, 
provide a family $\Phi(O^A)$ of functions which is rich enough to be 
able to approximate uniformly any spinor field by spinor fields of 
the structure $\pi_{A_1...A_{2s}}=\sum_{i=1}^N\phi^i\bar\chi^iC_i{}_{A
_1...A_{2s}}$. Here the combination $\phi^1\bar\chi^1+...+\phi^n\bar
\chi^n$ is an element of $\Phi(O^A)\otimes\bar\Phi(O^A)$, the tensor 
product of $\Phi(O^A)$ and its complex conjugate. However, $\bar\chi
(u,\bar\zeta)$ can also be interpreted as the profile function in the 
{\it pp}-wave solution of the {\it complex conjugate} Weyl equation 
$\nabla_A{}^{A'}\sigma_{A'}=0$. Its constant wave spinor is just the 
complex conjugate of $O^A$, and hence its wave vector is still 
that of the solution $\phi(u,\zeta)O^A$. Obviously, the two helicities 
of these solutions are just the opposite of each other. Therefore, 
the expansion that Theorem 5.2.1 provides is based on the {\it 
pp}-wave solutions both of the Weyl and the complex  conjugate Weyl 
equations with a given constant, common wave vector. Clearly, 
$\Phi(O^A)$ in itself with a single, fixed constant spinor field 
$O^A$ is {\it not} dense in $C^0(\Sigma,{\bf C})$, it is only the 
space $\Phi(O^A)\otimes\bar\Phi(O^A)$ of the homogeneous Hermitian 
quadratic expressions of the functions $\phi$ that is dense. 

Clearly, it is irrelevant which z.r.m. field equation (for strictly 
non-zero spin) is used to define the {\it pp}-waves: The structure 
of the {\it pp}-waves for any non-zero spin is similar. On the other 
hand, the {\it pp}-wave solutions of the {\it scalar} wave equation 
are plane waves, which for one given wave vector $L^a$ do {\it not} 
provide a big enough set $\Phi(L^a)$. Thus if we wanted to use only 
plane waves as a basis of approximation, then, by the 
Stone--Weierstrass theorem (see Appendix), we would have to take 
plane waves with {\it three independent} wave vectors $L^a_1$, $L^a_2$ 
and $L^a_3$, and it would be $\Phi(L^a_1)\otimes\Phi(L^a_2)\otimes\Phi
(L^a_3)$ whose density in $C^0(\Sigma,{\bf C})$ could be proven. The 
standard plane wave expansions in physics are such, and, as a 
consequence, the approximation is made by a {\it cubic} (rather than 
a quadratic) expression of the plane waves. The advantage of the 
approximation based on the {\it pp}-waves (rather than only plane 
waves) is that we should use solutions with {\it only one constant 
spinor field $O^A$ and its complex conjugate $\bar O^{A'}$}. Although 
$\Phi(O^A)$ with fixed $O^A$ is not dense in $C^0(\Sigma,{\bf C})$, the 
set of the {\it pp}-waves {\it with all the constant spinor fields} can 
be proven to be dense. Note that the approximation based on the 
genuine {\it pp}-waves is possible only quasi-locally: Both in the 
asymptotically flat and the closed cases, discussed at the 
beginning of this subsections, the {\it pp}-waves reduce to plane 
waves. 

As is well known, uniform convergence in a space of continuous 
functions can be characterized by convergence with respect to the 
$L_\infty$ (or supremum) norm. Since by Theorem 5.2.1 the space 
$C^0(\Sigma,{\bf S}_{(A_1...A_{2s})})$ of the continuous, totally 
symmetric spinor fields on $\Sigma$ is closed with respect to the 
uniform closure, it is complete with respect to the norm $\Vert
\psi_{A_1...A_{2s}}\Vert_\infty:=\sup_{p\in\Sigma}\vert\psi_{A_1...A
_{2s}}(p)\vert$, where $\vert\,.\,\vert$ is the pointwise Hermitian 
norm used in Theorem 5.2.1. However, by the compactness of $\Sigma$, 
$C^0(\Sigma,{\bf S}_{(A_1...A_{2s})})\subset\Gamma_s$ is a subspace and, 
as a simple consequence of the definitions, $\Vert\psi_{A_1...A_{2s}}
\Vert\leq\sqrt{{\rm Vol}(\Sigma)}\Vert\psi_{A_1...A_{2s}}\Vert_\infty$ 
holds, where ${\rm Vol}(\Sigma)$ is the metric volume of $\Sigma$. 
Consequently, Theorem 5.2.1 implies that any square-integrable 
totally symmetric spinor field can also be approximated by the 
spinor fields $\pi_{A_1...A_{2s}}$ above with arbitrary accuracy in 
the $L_2$ norm. Or, in other words, {\it the spinor fields $\pi
_{A_1...A_{2s}}$ built from the {\it pp}-wave solutions of the Weyl 
and the complex conjugate Weyl equations even with fixed constant 
spinor field $O^A$ form a dense subspace in $\Gamma_s$}. 

The spinor field $\psi_{A_1...A_{2s}}$ in Theorem 5.2.1 is not 
required to satisfy any field equation. If, for example, it is a 
solution of the linear constraint equation (4.1.2), then Theorem 
5.2.1 implies that the spinor fields of the form $\pi_{A_1...A_{2s}}$ 
above form a dense subspace in the `constraint surface' $\hat\Gamma
_s\subset\Gamma_s$. 
An even more interesting case is when $\phi^\alpha_{AB}$ is the 
spinor form of the Yang--Mills field strength with a non-Abelian 
gauge group, and hence satisfies a {\it nonlinear} spinor field 
equation on $\Sigma$. Then the set of the solutions is a subset of 
$\Gamma_1\times...\times\Gamma_1$ (${\rm dim}\,{\cal G}$-times), in 
which the spinor fields $\pi^\alpha_{AB}$ form a dense subset. Thus 
the solutions of {\it nonlinear} field equations on $\Sigma$ can be 
uniformly approximated by spinor fields built from the {\it pp}-wave 
solutions of {\it linear} equations. 

The fact that the algebra $\Phi(O^A)$ is not only a class of functions 
that separates points of $\Sigma$ but it is essentially the set of the 
{\it pp}-wave {\it solutions} is crucial from the point of view of 
holography. We saw in subsection 5.1 that each {\it pp}-wave solution
can be characterized completely by the holographic data on ${\cal S}
=\partial\Sigma$. Since by Theorem 5.2.1 every solution of the field 
equations can be approximated by finitely many {\it pp}-waves with 
arbitrary accuracy, it follows that every solution can also be 
approximated by the holographic data for finitely many {\it pp}-waves 
as well. Hence the `screen map' $\Sigma\rightarrow{\cal S}$ of the 
Introduction is replaced by $\Phi(O^A)\rightarrow{\rm `the \,\,set\,\, 
of \,\,the \,\,holographic \,\,data \,\,on \,\,{\cal S}'}$, providing a 
rather non-local representation of the fields on $\Sigma$ in terms of 
fields on ${\cal S}$. Thus $\Phi(O^A)$ is similar to a spider's web, 
where the space $\Sigma$ is `scanned' by the {\it pp}-waves and the 
information gained is encoded into the holographic data on ${\cal S}$. 
Finally, if we fix a foliation $\{\Sigma_t\}$ of $D(\Sigma)$ by smooth 
spacelike hypersurfaces, then by mapping the fields induced on each 
of the leaves $\Sigma_t$ from $D(\Sigma)$ to the corresponding 
holographic data on ${\cal S}$, we can represent every continuous 
spinor field on $D(\Sigma)$ by a 1-parameter family of holographic 
data. Therefore, quasi-local holography for the classical fields 
considered here works in Minkowski spacetime. 

\bigskip

\ni
{\lbf Appendix: The Stone--Weierstrass theorem}
\bigskip
\ni
The form of the Stone--Weierstrass theorem that motivated 
Theorem 5.2.1, quoted from [23], is 
\medskip
\ni
{\bf Theorem:} 
Let $X$ be a compact Hausdorff space and $f:X\rightarrow{\bf R}$ be 
continuous. If $\Phi$ is a set of continuous functions $\phi:
X\rightarrow{\bf R}$ such that $\Phi$ separates the points of $X$ 
(in the sense that for any two different points $p,q\in X$ there 
exists a function $\phi\in\Phi$ such that $\phi(p)\not=\phi(q)$), 
then for any $\varepsilon>$ there exist finitely many functions 
$\phi^1,...,\phi^k\in\Phi$ and a real polynomial $P=P(x^1,...,x^k)$ 
with $k$ variables such that $\vert f(p)-P(\phi^1(p),...,\phi^k(p))
\vert<\varepsilon$ for all $p\in X$. 
\medskip
\ni
Thus the role of $\Phi$ is to provide a `coordinate system' on the 
space of continuous real functions on $X$, and if it is `rich 
enough' then the elements of the unital algebra generated by $\Phi$ 
is uniformly dense in the space $C^0(X,{\bf R})$ of the continuous 
real functions endowed with the supremum norm. 

The classical Weierstrass approximation theorem that we used in 
the proof of Theorem 5.2.1 is its simple consequence: If $f:[a,b]
\subset{\bf R}\rightarrow{\bf R}$ is continuous, then for any 
$\varepsilon>0$ there exists a polynomial $P$ such that $\vert 
f(x)-P(x)\vert<\varepsilon$ for all $x\in[a,b]$. 

\bigskip
\ni
{\lbf Acknowledgments}
\bigskip
\ni
The author is grateful to J\"org Frauendiener for the exhaustive 
discussion on the solutions of the linear z.r.m. field equations 
and their twistor theoretical aspects, and to Zolt\'an Perj\'es 
for his remarks on the elementary states. This work was partially 
supported by the Hungarian Scientific Research Fund grant OTKA 
T042531. 

\bigskip
\ni
{\lbf References}
\bigskip

\item{[1]} G. 't Hooft, Dimensional reduction in quantum gravity, 
           gr-qc/9310026
\item{[2]} J.D. Bekenstein, A universal upper bound on the entropy 
           to energy ratio for bounded systems, Phys. Rev. D {\bf 23} 
           287--298 (1981)
\item{[3]} R. Bousso, A covariant entropy conjecture, JHEP 07 
           (1999) 004, hep-th/9905177v3
\item{[4]} L. Susskind, The world as a hologram, J. Math. Phys. 
           {\bf 36} 6377--6396 (1995), hep-th/9409089
\item{[5]} S. Corley, T. Jacobson, Focusing and the holographic 
           hypothesis, Phys. Rev. D {\bf 53} R6720--R6724 (1996), 
           gr-qc/9602043
\item{[6]} R. Bousso, The holographic principle, Rev. Mod. Phys. 
           {\bf 74} 825--874 (2002), hep-th/0203101v1 

\item{[7]} L.B. Szabados, Quasi-local energy-momentum and angular 
           momentum in GR: A review article, Living Rev. Relativity 
           {\bf 7} (2004) 4. [Online article]: {\tt 
           http://www.livingreviews.org/lrr-2004-4 }

\item{[8]} A.J. Dougan, L.J. Mason, Quasilocal mass constructions 
           with positive energy, Phys. Rev. Lett. {\bf 67} 2119--2122 
           (1991)
\item{[9]} L.B. Szabados, On the positivity of the quasi-local mass, 
            Class. Quantum Grav. {\bf 10} 1899--1905 (1993) 
\item{[10]} L.B. Szabados, Two dimensional Sen connections and 
            quasi-local energy-momentum, Class. Quantum Grav, {\bf 
             11} 1847--1866 (1994), gr-qc/9402005
\item{[11]} L.B. Szabados, Quasi-local energy-momentum and 
            two-surface characterization of the {\it pp}-wave 
            spacetimes, Class. Quantum Grav. {\bf 13} 1661--1678 
            (1996), gr-qc/9512013

\item{[12]} R. Penrose, W. Rindler, {\it Spinors and Space-time}, Vol 
           1, Two-Spinor Calculus and Relativistic Fields, 
           Cambridge Univ. Press, Cambridge 1984

\item{[13]} K.P. Tod, The St\"utzfunktion and the cut function, in 
           {\it Recent Advances in General Relativity: Essays in honor 
           of Ezra Newman}, Vol 4 of Einstein studies, pp. 182--195, 
           Eds.: A.I. Janis and J.R. Porter, Birkh\"auser, Boston 
           1992

\item{[14]} S. Coleman, Non-Abelian plane waves, Phys. Lett. {\bf 
           70B} 59--60 (1977)
\item{[15]} R. G\"uven, Solution for gravity coupled to non-Abelian 
           plane waves, Phys. Rev. D {\bf 19} 471--472 (1979) 

\item{[16]} J. Frauendiener, G.A.J. Sparling, On a class of consistent 
           linear higher spin equations on curved manifolds, J. Geom. 
           Phys. {\bf 30} 54--101 (1999) 
\item{[17]} S.A. Hugget, K.P. Tod, {\it An Introduction to Twistor 
          Theory}, London Mathematical Society Student Texts No 4, 
          Cambridge University Press, Cambridge 1985

\item{[18]} A. Sen, On the existence of neutrino `zero-modes' in vacuum 
           spacetimes, J. Math. Phys. {\bf 22} 1781--1786 (1981) 
\item{[19]} R.M. Wald, {\it General Relativity}, Univ. Chicago Press, 
           Chicago 1984

\item{[20]} R. Penrose, Solutions of the zero-rest-mass equations, 
           J. Math. Phys. {\bf 10} 38--39 (1969)  
\item{[21]} R. Penrose, M.A.H. MacCallum, Twistor theory: An approach 
          to the quantisation of fields and space-time, Phys. Rep. 
          {\bf 6} 241--316 (1972)
\item{[22]} R. Penrose, W. Rindler, {\it Spinors and Space-time}, Vol 
           2, Spinor and Twistor Methods in Space-time Geometry, 
           Cambridge Univ. Press, Cambridge 1986 
\item{[23]} M.G. Eastwood, A.M. Pilato, On the density of twistor 
           elementary states, Pacific J. Math. {\bf 151} 201--215 
           (1991) 

\item{[24]} B. Sz.-Nagy, {\it Introduction to Real Functions and 
           Orthogonal Expansion}, Joint edition published by 
           Akad\'emiai Kiad\'o, Budapest and Oxford University 
           Press, New York 1964 
\item{[25]} G.K. Pedersen, {\it Analysis Now}, Graduate Texts in 
           Mathematics Vol 118, Springer--Verlag, New York 1989

\end